\documentclass[journal]{IEEEtran}

\usepackage{graphicx}
\usepackage{amsfonts}
\usepackage{amsmath}
\usepackage{amssymb}
\usepackage{color}

\newtheorem{thm}{Theorem}[section]
\newtheorem{cor}[thm]{Corollary}
\newtheorem{lem}[thm]{Lemma}
\newtheorem{proof}[thm]{proof}

\newtheorem{defn}[thm]{Definition}
\newtheorem{rem}[thm]{Remark}
\newtheorem{exam}[thm]{Example}

\numberwithin{equation}{section}

\ifCLASSINFOpdf \else \fi 
\begin{document}

\title{The Complexity of Network Coding with \\Two Unit-Rate Multicast Sessions}

\author{Wentu Song, ~  Kai Cai, ~  Rongquan Feng ~  and ~ Chau Yuen
\thanks{Wentu Song is with the LMAM, School of Mathematical Sciences,
Peking University, China, and with Singapore University of
Technology and Design, Singapore
       (e-mails: songwentu@gmail.com).}
\thanks{Kai Cai is with the Institute of Network Coding, The Chinese University
of Hong Kong, Shatin, New Territories, Hong Kong. (e-mail:
eecaikai@gmail.com).}
\thanks{Rongquan Feng is with the LMAM, School of Mathematical Sciences,
Peking University, China
       (e-mails: fengrq@math.pku.edu.cn).}
\thanks{Chau Yuen is with Singapore University of Technology and Design,
Singapore (email: yuenchau@sutd.edu.sg).}
\thanks{Copyright (c) 2012 IEEE. Personal use of this material is permitted.
However, permission to use this material for any other purposes
must be obtained from the IEEE by sending a request to
pubs-permissions@ieee.org.}
}

\markboth{Journal of \LaTeX\ Class Files,~Vol.~11, No.~4, December~2012}%
{Shell \MakeLowercase{\textit{et al.}}: Bare Demo of IEEEtran.cls
for Journals}

\maketitle

\begin{abstract}
The encoding complexity of network coding for single multicast
networks has been intensively studied from several aspects: e.g.,
the time complexity, the required number of encoding links, and
the required field size for a linear code solution. However, these
issues as well as the solvability are less understood for networks
with multiple multicast sessions. Recently, Wang and Shroff showed
that the solvability of networks with two unit-rate multicast
sessions (2-URMS) can be decided in polynomial time
\cite{Wang110}. In this paper, we prove that for the 2-URMS
networks: $1)$ the solvability can be determined with time
$O(|E|)$; $2)$ a solution can be constructed with time $O(|E|)$;
$3)$ an optimal solution can be obtained in polynomial time; $4)$
the number of encoding links required to achieve a solution is
upper-bounded by $\max\{3,2N-2\}$; and $5)$ the field size
required to achieve a linear solution is upper-bounded by
$\max\{2,\lfloor\sqrt{2N-7/4}+1/2\rfloor\}$, where $|E|$ is the
number of links and $N$ is the number of sinks of the underlying
network. Both bounds are shown to be tight.
\end{abstract}

\begin{IEEEkeywords}
Network coding, encoding complexity, region decomposition.
\end{IEEEkeywords}


\section{Introduction}
\IEEEPARstart{A} communication network is described as a finite,
directed, acyclic graph $G=(V,E)$, where a number of messages are
generated at some nodes, named {\em sources}, and are demanded by
some other nodes, named {\em sinks}. Messages are assumed to be
independent random processes with the values taken from some fixed
finite alphabet, usually a finite field. Network coding allows the
intermediate nodes to ``encode'' the received messages before
forwarding it, and has significant throughput advantages as
opposed to the conventional store-and-forward scheme
\cite{Ahlswede00, Li03}. In literature, most research of network
coding focused on multicast networks. For nonmulticast networks,
there are only a few results, for example, some deterministic
results on the capacity region for some specific networks, such as
single-source two-sink nonmulticast networks \cite{Nga04},
directed cycles \cite{Harv06}, degree-2 three-layer directed
acyclic networks \cite{Yan06}, and networks with two unit-rate
multicast sessions (2-URMS) \cite{Wang110}. Some outer bounds on
the capacity region for general nonmulticast networks were
obtained by information theoretic arguments
\cite{Harv06}-\cite{Yan07} and some inner bounds were obtained by
linear programming \cite{Tra06, Wu06}. In \cite{Lehman03} it was
proved that determining whether there exist linear network coding
solutions for an arbitrary nonmulticast network is NP-hard.

Besides the solvability, another important issue of network coding
problem is the encoding complexity. The encoding complexity of
multicast networks has been intensively studied
\cite{Jaggi05}-\cite{Fragouli06}. However, for nonmulticast
networks, it remains challenging due to the intrinsic hardness of
the nonmulticast network coding problem. In previous works, the
encoding complexity is generally studied from three aspects: the
time complexity for constructing a network coding solution, the
number of the required encoding nodes, and the required field size
for achieving a network coding solution.

The time complexity is a fundamental issue of network coding
complexity. It is well known that a network coding solution can be
achieved with polynomial time for multicast networks
\cite{Jaggi05}. In \cite{Langberg06}, the authors first
categorized the network links into two classes, i.e., the {\em
forwarding links} and the {\em encoding links}. The forwarding
links only forward the data received from its incoming links.
While, the encoding links transmit {\em coded packets}, which need
more resources due to the computing process and encoding
capabilities. It was shown that, in an acyclic multicast network,
the number of encoding nodes (i.e., the tail of an encoding link)
required to achieve the capacity of the network is independent to
the size of the underlying network and is bounded by $h^3N^2$,
where $N$ is the number of the sinks and $h$ is the number of the
source messages. The third aspect of encoding complexity is the
required field size. As mentioned in \cite{Dougherty04}, a larger
encoding field size may cause difficulties, i.e. either a larger
delay or a larger bandwidth for the implementation of network
coding. Hence, smaller alphabets are preferred. For the multicast
network, the required alphabet size to achieve a solution is upper
bounded by $N$ (see \cite{Jaggi05}).

In \cite{Fragouli06}, a method called information flow
decomposition was proposed, which is efficient to decrease the
complexity of network code design for single session multicast
networks. For a solvable single session multicast network with $h$
unit rate sources and $N$ sinks (receivers), we can find $h$
edge-disjoint paths from the sources to each sink. By the
information flow decomposition approach, the line graph of the
$hN$ paths is decomposed into subtrees. When performing network
coding, the same coding vector is assigned to all links in the
same subtree and the network can be contracted to a subtree graph.
The subtree graph can be efficiently reduced to a minimal subtree
graph while maintaining the ``multicast property''. The optimal
network code solution can be obtained from the code on the minimal
subtree graph. Note that the choice of these $hN$ paths is not
unique, and will affect the complexity of the network code. The
authors showed that for multicast networks with two unit-rate
sources, the number of coding subtrees is not greater than $N-1$
and a finite field with size $\sqrt{2N-7/4}+1/2$ is sufficient to
achieve a linear solution.

In this paper, we consider the encoding complexity of networks
with two unit-rate multicast sessions (2-URMS), where two message
symbols are generated by two sources and are demanded by two sets
of sinks respectively. If the two sink sets are identical, it is a
multicast network coding problem, of which the solvability can be
characterized by the well-known max-flow condition and its
encoding complexity has been discussed as mentioned above.
However, in the case the two sink sets are distinctly different,
the situation becomes complicated. The recent work of Wang and
Shroff \cite{Wang110} showed that the solvability of the 2-URMS
problem can be characterized by \emph{paths with controlled
edge-overlap} condition under the assumption of sufficiently large
encoding fields. They also proved that deciding the solvability of
a 2-URMS network is polynomial time complexity and linear network
codes are sufficient to achieve a solution. In addition, there are
also some works that considering different scenarios, e.g., the
2-unicast setting, for which the reader can refer to
\cite{Wang11}, \cite{Shomorony11}, \cite{Kamath11}, \cite{Cai11}.

We further developed the information flow decomposition approach
\cite{Fragouli06} and proposed a region decomposition method for
network coding of 2-URMS networks by which the whole network is
decomposed into mutually disjoint {\em regions}. As in
\cite{Fragouli06}, we assign the same coding vector to all links
in the same region and contract the network to a {\em region
graph} when performing network coding. A region graph is said to
be feasible if there is a network coding solution to it. We prove
that each network has a unique region decomposition, called the
basic region decomposition, and the solvability of the network is
equivalent to the feasibility of the basic region graph (i.e., the
region graph corresponding to the basic region decomposition). We
give a sufficient and necessary condition for feasibility of
region graph of 2-URMS networks which can be verified in time
$O(|E|)$, where $|E|$ is the number of links of the network. If a
network is solvable, we can construct a linear solution on the
basic region graph in a decentralized manner, which significantly
reduces the complexity of network code design compared to the
method in \cite{Wang110}. Moreover, any feasible region graph can
be reduced to a minimal feasible region graph and an optimal
solution can be obtained from network code on the minimal feasible
region graph.

The main differences between the region decomposition in this paper
and the subtree decomposition approach in \cite{Fragouli06} are as
follows: (1) As stated before, subtree decomposition is performed on
a subnetwork of the original network, the choice of which is not
unique; while region decomposition is performed on the whole network
and for each network, we can obtain a unique basic region
decomposition and the solvability of the original network can be
characterized by the feasibility of the basic region graph. (2) A
region in our paper need not be a tree, in fact it can be any kind
of graph configuration. (3) Unlike the subtree graph in
\cite{Fragouli06}, we cannot use the multicast property to determine
feasibility of region graph of 2-URMS networks, since the
max-flow/min-cut condition is not sufficient to characterize the
solvability of non-multicast networks \cite{Koetter03}. For this
reason, we give a new characterization of feasibility for region
graph of 2-URMS networks based on region labelling. (4) In
\cite{Fragouli06}, each sink (receiver) is contained in $h$
subtrees, where $h$ is the source rate, and all subtrees are divided
into two classes, i.e., source subtree and coding subtree. In this
paper, each sink is contained in exactly one region, named as sink
region. Hence, different from \cite{Fragouli06}, besides source
region and coding region, there are sink regions in this paper. In
addition, for solvable single session multicast networks, the
max-flow between the sources and the sinks equals to the source
rate; while for 2-URMS networks, the max-flow between the sources
and some sinks can be only one. So it is necessary to introduce the
new class of sink region. The presence of sink regions increases the
hardness of network code design on region graph.

The main contributions of this paper are listed as below:
\begin{itemize}
\item We characterize the solvability of a 2-URMS network, which
can be verified in time $O(|E|)$, and we design network code with
$O(|E|)$ time algorithm for a solvable 2-URMS network, where $|E|$
is the number of the links of the underlying network. \item  We
present a polynomial time algorithm to construct the {\em optimal}
solutions for solvable 2-URMS networks, where an optimal solution
means a solution with a minimum number of encoding links. \item We
prove that the number of encoding links for achieving a solution
is upper-bounded by $\max\{3,2N-2\}$, where $N$ is the number of
sinks. Note that it is independent of the network size and is only
related to $N$. We also give instances which achieve this bound.
\item We prove that the required field size to achieve a linear
solution of a 2-URMS network is upper-bounded by
$\max\{2,\lfloor\sqrt{2N-7/4}+1/2\rfloor\}$, which is {\em
surprisingly as small as the multicast case}. Also, this bound is
shown to be tight by constructing some instances.
\end{itemize}

The rest of this paper is organized as follows. In Section
\uppercase\expandafter{\romannumeral 2}, we give the network model
and some basic definitions of network coding. In Section
\uppercase\expandafter{\romannumeral 3}, we introduce the method
of region decomposition, including the definitions of region,
region decomposition, region graph, region contraction, codes on
the region graph, feasible region graph, region labeling, etc. We
also derive some basic properties for these notions in this
section. In Section \uppercase\expandafter{\romannumeral 4}, we
decide the time complexity for solving the 2-URMS problem by
introducing the basic region decomposition. We introduce the
minimal feasible region graph and investigate the optimal solution
in Sections \uppercase\expandafter{\romannumeral 5}, the number of
required encoding links is given in the same section. The required
encoding field size is obtained in Section
\uppercase\expandafter{\romannumeral 6}. Finally, we conclude the
paper in Section \uppercase\expandafter{\romannumeral 7}.

\section{Network Model and Notations}
We consider the network coding problem with two unit-rate
multicast sessions (2-URMS), of which the underlying network is a
finite, directed, acyclic graph $G=(V,E)$, where $V$ is the set of
nodes (vertices) and $E$ is the set of links (directed edges).
There are two sources $s_1,s_2\in V$ and two sets of sinks
$T_1=\{t_{1,1},\cdots,t_{1,N_1}\}$,
$T_2=\{t_{2,1},\cdots,t_{2,N_2}\}\subseteq V$, where $s_i\notin
T_i (i=1,2)$. Two messages $X_1$ and $X_2$ are generated by $s_1$
and $s_2$ and are demanded by sinks in $T_1$ and $T_2$
respectively. Note that $T_1\neq T_2$ in general. The messages are
assumed to be independent random variables taking values from a
fixed finite field and a link $e=(u,v)$ is assumed of unit
capacity, i.e., it can transmit one message symbol from node $u$
to node $v$ per transmission.

For any link $e=(u,v)\in E$, $e$ is called an outgoing link of $u$
and an incoming link of $v$. Meanwhile, $u$ is called the tail of
$e$ and $v$ is called the head of $e$ and denoted as
$u=\text{tail}(e)$ and $v=\text{head}(e)$. For $e_1,e_2\in E$, we
call $e_1$ an {\em incoming link} of $e_2$ if
$\text{head}(e_{1})=\text{tail}(e_{2})$. And $\text{In}(e)$ denotes
the set of the incoming links of $e$.

We assume that the source $s_i$ has an imaginary incoming link,
called the $X_{i}$ {\em source link} (or \emph{source link} for
short), and each sink $t_{i,j}\in T_{i}$ has an imaginary outgoing
link, called the $X_{i}$ {\em sink link} (or \emph{sink link} for
short). Both $X_{1}~($resp. $X_2)$ source link and $X_{1}~($resp.
$X_2)$ sink link are called $X_{1}~($resp. $X_2)$ link. Note that
the source links have no tail and the sink links have no head. As
a result, the source links have no incoming link. For the sake of
convenience, if $e\in E$ is not a source link, we call $e$ a
non-source link.

We assume that $\text{In}(e)\neq\emptyset$ for each non-source
link $e\in E$. Otherwise $e$ has no impact on the network
communication and can be removed from $G$.

\begin{rem}\label{indexE}
Since $G=(V,E)$ is an acyclic graph, $E$ can be sequentially
indexed as $e_1,e_2,e_3,\cdots,e_{|E|}$ such that: 1) $e_1$ is the
$X_1$ source link and $e_2$ is the $X_2$ source link; 2) $i<j$ if
$e_i$ is an incoming link of $e_j$.
\end{rem}

Following \cite{Yeung06}, the network coding solution of a 2-URMS
network is defined as follows.

\begin{defn}[Network Coding Solution]\label{LSo}
A \emph{network coding solution} (or \emph{a solution} for short)
of $G$ over the field $\mathbb F$ is a collection of functions
$C=\{f_{e}: \mathbb F^2\rightarrow \mathbb F; e\in E\}$ such that
\begin{itemize}
    \item[(1)] If $e$ is an $X_{i}$ link,
    then $f_{e}(X_1,X_2)=X_{i}~(i\in\{1,2\})$.
    \item[(2)] If $e$ is a non-source link, then $f_e$ can be computed from
    $f_{p_{1}},\cdots,f_{p_{k}}$, where $\{p_{1},\cdots,p_{k}\}=\text{In}(e)$.
    This means that there is a $\mu_e: \mathbb F^{k}\rightarrow \mathbb F$
    such that $f_{e}=\mu_e(f_{p_{1}},\cdots,f_{p_{k}})$.
\end{itemize}
\end{defn}
The function $f_e$ is called the {\em global encoding function} of
$e$ and $\mu_e$ is called the {\em local encoding function} of
$e$. A solution $C$ is called a {\em linear solution} if the local
encoding functions are all linear functions over $\mathbb F$.

A network $G$ is said to be {\em (linearly) solvable} if $G$ has a
(linear) solution over some finite field.

\begin{rem}\label{RLSo}
As in \cite{Yeung06}, a linear network coding solution of $G$ over
the field $\mathbb F$ can be described as a collection of vectors
$C=\{d_{e}\in \mathbb F^{2}; e\in E\}$ such that the following two
conditions hold:
\begin{itemize}
 \item[($1'$)] If $e$ is an $X_{i}$ link, then
 $d_{e}=\alpha_{i}~(i\in\{1,2\})$, where $\alpha_1=(1,0)$ and $\alpha_2=(0,1)$;
 \item[($2'$)] If $e$ is a non-source link, then $d_{e}$ is an $\mathbb F$-linear
 combination of $\{d_{p};p\in\text{In}(e)\}$.
\end{itemize}
In this case, $d_e$ is called the {\em (global) coding vector} of
$e$.
\end{rem}

\begin{defn}[Forwarding Link and Encoding Link]\label{EEFE}
Let $C=\{f_{e}; e\in E\}$ be a solution of $G$. A link $e$ is
called a \emph{forwarding link} of $C$ if $f_e=f_p$ for some
$p\in\text{In}(e)$. Else, $e$ is called an \emph{encoding link} of
$C$.
\end{defn}

Note that there are works that classify the operations at the node
level, e.g., \cite{Bhattad05}, however, in this paper, we adopt
link level classification. As pointed out in \cite{Langberg06},
the encoding links is an accurate estimation of the total amount
of computations performed by a coding network and is more closely
related to the encoding complexity.

We will need to use the line graph of a network $G=(V,E)$, denoted
by $L(G)$, which is defined as a directed, simple graph with vertex
set $E$ and edge set $\{(e_i,e_j)\in E^2; e_i$ is an incoming link
of $e_j\}~($e.g., see \cite{Koetter03}$)$. The line graph $L(G)$ is
obviously finite and acyclic, since $G$ is finite and acyclic.

\section{Region Decomposition}
In this section, we introduce the region decomposition method and
investigate the code construction on the region graph. The basic
idea of region decomposition is the same as the subtree
decomposition \cite{Fragouli06}. Specifically, we decompose the
network into mutually disjoint {\em regions} and assign the same
coding vector to all links in the same region for the network code
design problem. By doing so, the structure inside of a region does
not play any role; we only need to know how the regions are
connected. Thus, we can contract each region to a node and retain
only the edges that connect them. We will demonstrate that such
decomposition of the network can significantly decrease the
encoding complexity.

\subsection{Region Decomposition and Region Graph}
Most definitions of region decomposition are analogous to the
corresponding definitions of subtree decomposition. We first
define region decomposition and region graph of a network, which
are analogue to the subtree decomposition and subtree graph in
\cite{Fragouli06} respectively.

\begin{defn}[Region and Region Decomposition]\label{Reg}
Let $R$ be a non-empty subset of $E$. $R$ is called a region of
$G$ if there is an $e_{l}\in R$ such that for any $e\in R$ and
$e\neq e_{l}$, $R$ contains an incoming link of $e$. If $E$ is
partitioned into mutually disjoint regions, say
$R_{1},R_{2},\cdots,R_{n}$, then we call
$D=\{R_{1},R_{2},\cdots,R_{n}\}$ a region decomposition of $G$.
\end{defn}

The edge $e_{l}$ in Definition \ref{Reg} is called the leader of
$R$ and is denoted as $e_{l}=\text{lead}(R)$. Since $G$ is
acyclic, it is easy to see that the leader of a region is unique.
Let $D$ be a region decomposition of $G$ and $R\in D$. $R$ is
called the $X_{1}~($resp. $X_2)$ \emph{source region} if
$\text{lead}(R)$ is the $X_{1}~($resp. $X_2)$ source link; $R$ is
called an $X_{1}~($resp. $X_2)$ \emph{sink region} if $R$ contains
an $X_{1}~($resp. $X_2)$ sink link. Both the $X_{1}$ source region
and $X_{2}$ source region are called {\em source region}, and both
the $X_{1}$ sink region and $X_{2}$ sink region are called {\em
sink region}. For the sake of convenience, if $R$ is not a source
region, we call $R$ a \emph{non-source region}.

A network may have many region decompositions. For example,
$\forall e\in E$, $R_e=\{e\}$ is a region with
$\text{lead}(R_e)=e$ and $D^*=\{R_e; e\in E\}$ is a region
decomposition of $G$. We call $D^*$ the {\em trivial region
decomposition} of $G$.

The following is an example of nontrivial region decomposition.
This example network will be used frequently in the sequel.

\begin{figure}[htbp]
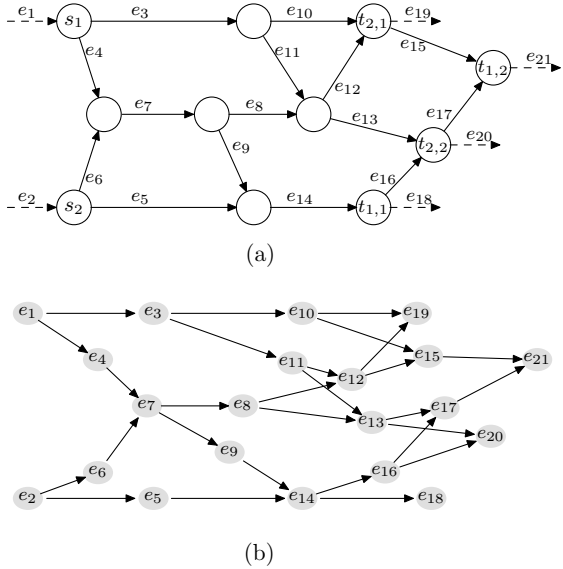

\begin{center}
\includegraphics[width=7.4cm]{redefig1.1}
\includegraphics[width=7.2cm]{redefig1.2}
\end{center}
\caption{(a) is an example network $G_1$, where messages $X_{1}$
and $X_{2}$ are generated at sources $s_{1}$ and $s_{2}$ and are
demanded by sinks in $T_1=\{t_{1,1},t_{1,2}\}$ and
$T_2=\{t_{2,1},t_{2,2}\}$ respectively. The imaginary edges
$e_{1}$ and $e_{2}$ are the $X_{1}$ source link and the $X_{2}$
source link respectively; the imaginary links $e_{18}, e_{21}$ are
$X_{1}$ sink links and the imaginary links $e_{19},e_{20}$ are
$X_2$ sink links. (b) is the line graph $L(G_1)$ of $G_1$. By
Definition \ref{RegG}, $L(G_1)$ is also the trivial region graph
$\text{RG}(D^*)$ of $G_1$, where $D^*$ is the trivial region
decomposition of $G_1$.} \label{EG1-f}
\end{figure}

\begin{exam}\label{ERG1}
Consider the network $G_1$ in Fig. \ref{EG1-f}(a). Let
$R_1=\{e_{1},e_{3},e_{4},e_{10},e_{11}\}$,
$R_2=\{e_{2},e_{5},e_{6}\}$,
$R_3=\{e_{7},e_{8},e_{9},e_{12},e_{13},e_{15}\}$,
$R_4=\{e_{14},e_{16},e_{18}\}$, $R_5=\{e_{17}\}$,
$R_6=\{e_{19}\}$, $R_7=\{e_{20}\}$ and $R_8=\{e_{21}\}$. Then
$D_1=\{R_1,\cdots,R_8\}$ is a region decomposition of $G_1$, which
is illustrated in Fig. 2(a). The region $R_1$ is the $X_1$ source
region since $\text{lead}(R_1)=e_1$ is the $X_1$ source link
$e_1$. Similarly, $R_2$ is the $X_2$ source region, $R_4$ and
$R_8$ are $X_1$ sink regions, $R_6$ and $R_7$ are $X_2$ sink
regions.
\end{exam}

\renewcommand\figurename{Fig}

Let $D$ be a region decomposition of $G$. We can define region
graph of $G$ as follows.

\begin{defn}\label{RegG}
The region graph of $G$ about $D$ is a directed, simple graph with
vertex set $D$ and edge set $\mathcal E_D$, where $\mathcal E_D$
is the set of all ordered pairs $(R',R)$ such that $R'$ contains
an incoming link of $\text{lead}(R)$. More generally, if $\mathcal
G$ is a graph with vertex set $D$ and edge set $\mathcal
E\subseteq \mathcal E_D$, then we call $\mathcal G=(D,\mathcal E)$
a \emph{region graph} of $G$ belonging to $D$.
\end{defn}

We use $\text{RG}(D)$ to denote the region graph of $G$ about $D$,
i.e., $\text{RG}(D)=(D,\mathcal E_D)$. Thus, any region graph
$\mathcal G=(D,\mathcal E)$ is a subgraph of $\text{RG}(D)$.

By Definition \ref{RegG}, $\text{RG}(D)$ is uniquely determined by
$G$ and $D$. Since $G$ is acyclic, it is easy to see that
$\text{RG}(D)$ is also acyclic, and hence any region graph
$\mathcal G=(D,\mathcal E)$ is acyclic. If $(R',R)$ is an edge of
$\mathcal G$, we call $R'$ a {\em parent} of $R$ ($R$ a {\em
child} of $R'$) in $\mathcal G$. Two regions $R'$ and $R$ are said
to be adjacent in $\mathcal G$ if $R'$ is a parent or a child of
$R$ in $\mathcal G$. We use $\text{In}_{\mathcal G}(R)$ to denote
the set of all parents of $R$ in $\mathcal G$. If $\mathcal
G=\text{RG}(D)$, we omit the subscript $\mathcal G$ and denote
$\text{In}_{\mathcal G}(R)$ as $\text{In}(R)$.

Since we assume that $\text{In}(e)\neq\emptyset$ for each
non-source link $e$, so $\text{In}(R)\neq\emptyset$ for each
non-source region $R$. Meanwhile, since the source links have no
incoming link, the source regions have no parent in any region
graph $\mathcal G$.

Clearly, the region graph $\text{RG}(D^*)$ is just the line graph
$L(G)$, where $D^*$ is the trivial region decomposition of $G$. We
call $\text{RG}(D^*)$ the trivial region graph of $G$. An example
is shown in Fig. \ref{EG1-f}(b).

\renewcommand\figurename{Fig}
\begin{figure}[htbp]
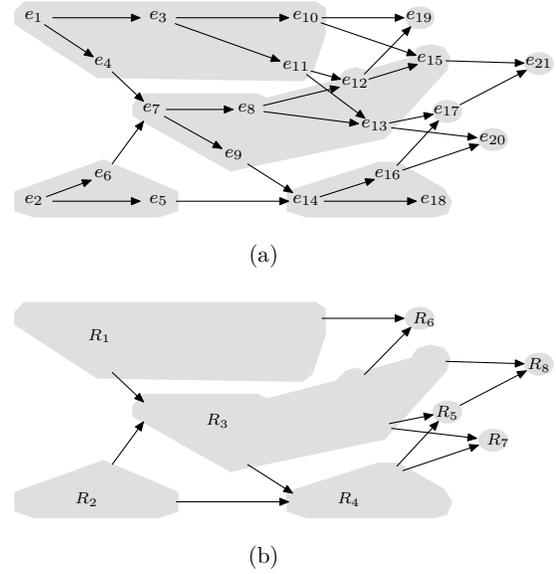

\begin{center}
\includegraphics[width=7.2cm]{redefig1.3}
\includegraphics[width=7.2cm]{redefig1.4}
\end{center}
\caption{Example of region decomposition and region graph: (a) The
region decomposition $D_1=\{R_1,\cdots,R_8\}$ of $G_1$, where
$G_1$ and $D_1$ are as in Example \ref{ERG1}. (b) The
corresponding region graph $RG(D_1)$.}\label{RGG1}
\end{figure}

For the example network $G_1$ and the region decomposition
$D_1~($See Fig. 1(a) and Fig. 2(a) respectively$)$, we demonstrate
the region graph $\text{RG}(D_1)$ in Fig. \ref{RGG1}(b). By
Definition \ref{RegG}, deleting any edge(s) of $\text{RG}(D_1)$ we
can obtain a region graph of $G$ belonging to $D_1$.

\begin{lem}\label{CFR}
Suppose $D$ is a region decomposition of $G$ and $\{P,Q\}\subseteq
D$ such that $P$ is a parent of $Q$ in $\text{RG}(D)$. Then
$P'=P\cup Q$ is a region of $G$ with
$\text{lead}(P')=\text{lead}(P)$ and
$D'=D\cup\{P'\}\setminus\{P,Q\}$ is a region decomposition of $G$.
\end{lem}
\begin{proof}
The conclusion can be directly verified by Definition \ref{Reg}
and \ref{RegG}.
\end{proof}

\begin{defn}\label{ConR}
Under the conditions of Lemma \ref{CFR}, $D'$ is called a
contraction of $D$ by combining $P$ and $Q$. Correspondingly, the
region graph $\text{RG}(D')$ is called a contraction of
$\text{RG}(D)$ by combining $P$ and $Q$.
\end{defn}

More generally, if $\mathcal G=(D,\mathcal E)$ is a region graph
of $G$ belonging to $D$ and $D'$ is the contraction of $D$ by
combining two regions $P,Q\in D$ to a new region $P'\in D'$, we
can define a contraction $\mathcal G'$ of $\mathcal G$ as follows.

\begin{defn}\label{ConRG}
Let $\mathcal G'$ be the graph with vertex set $D'$ and edge set
$\mathcal E'=\mathcal E'_1\cup\mathcal E'_2\cup\mathcal E'_3$,
where $\mathcal E'_1=\{(R,P'); (R,P)\in\mathcal E\}$, $\mathcal
E'_2=\{(P',R); R\in D\setminus\{P,Q\}, (P,R)\in\mathcal E$ or
$(Q,R)\in\mathcal E\}$ and $\mathcal E'_3=\{(R',R);
\{R',R\}\subseteq D\setminus\{P,Q\}$ and $(R',R)\in\mathcal E\}$.
Then $\mathcal G'$ is called a contraction of $\mathcal G$ by
combining $P$ and $Q$.
\end{defn}

It is easy to see that $\mathcal E'\subseteq\mathcal E_{D'}$. So
$\mathcal G'$ is a region graph of $G$ belonging to $D'$.

\begin{figure}[htbp]
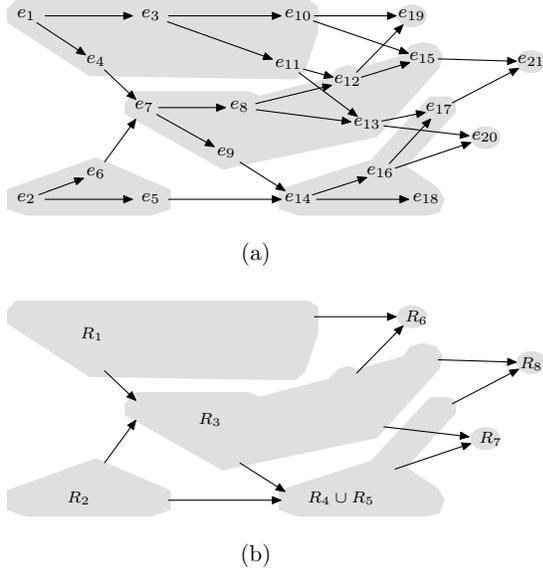

\begin{center}
\includegraphics[width=7.2cm]{redefig1.5}
\includegraphics[width=7.2cm]{redefig1.6}
\end{center}
\caption{Example of region contraction: (a) The region
decomposition $D_2$ of $G_1$, which is the contraction of $D_1$ by
combining $R_4$ and $R_5$. (b) The region graph $RG(D_2)$, where
$G_1$ and $D_1$ are shown in Fig. 1(a) and Fig. 2(a) respectively
(also in Example \ref{ERG1}).}\label{ECR1}
\end{figure}

\begin{exam}\label{EG1}
Consider the network $G_1$ and the region decomposition $D_1~($See
Fig. 1(a) and Fig. 2(a) respectively$)$. Clearly, $R_4\cup
R_5=\{e_{14},e_{16},e_{17},e_{18}\}$ is still a region of $G_1$
and $D_2=\{R_1,R_2,R_3,R_4\cup R_5,R_6,R_7,R_8\}$ is the
contraction of $D_1$ by combining $R_4$ and $R_5$. The region
decomposition $D_2$ and the region graph $\text{RG}(D_2)$ are
illustrated in Fig. \ref{ECR1}.
\end{exam}
\renewcommand\figurename{Fig}

Clearly, the edge set of $\text{RG}(D_2)$, i.e. $\mathcal
E_{D_2}$, can also be obtained as follows: Since $R_2$ and $R_3$
are parents of $R_4$ in $\text{RG}(D_1)$, then they are parents of
$R_4\cup R_5$ in $\text{RG}(D_2)$; Since $R_4$ is a parent of
$R_7$ and $R_5$ is a parent of $R_8$ in $\text{RG}(D_1)$, then
$R_4\cup R_5$ is a parent of $R_7$ and $R_8$ in $\text{RG}(D_2)$;
The rest edges of $\text{RG}(D_2)$ is the same as they are in
$\text{RG}(D_1)$.

By Definition \ref{ConR} and \ref{ConRG}, any region decomposition
of $G$ can be obtained from the trivial region decomposition of
$G$ by a sequence of region contraction. Correspondingly, any
region graph of $G$ can be obtained from the trivial region graph
of $G$ by a sequence of region contraction.

\subsection{Codes on the Region Graph}
In this subsection, we investigate code construction on region
graph.

\begin{defn}[Codes On Region Graph]\label{Fea}
Suppose $D$ is a region decomposition of $G$ and $\mathcal
G=(D,\mathcal E)$ is a region graph of $G$ belonging to $D$.
Suppose $\tilde{C}=\{f_{R}: \mathbb F^2\rightarrow \mathbb F; R\in
D\}$ is a collection of functions. Then $\tilde{C}$ is said to be
a code of $\mathcal G$ if the following two conditions hold:
\begin{itemize}
    \item[(1)] If $R$ is an $X_{i}$ source region or an $X_{i}$ sink region,
    then $f_{R}(X_1,X_2)=X_{i}~(i\in\{1,2\})$;
    \item[(2)] If $R$ is a non-source region, then $f_R$ can be computed from
    $f_{P_{1}},\cdots,f_{P_{k}}$, where $\{P_{1},\cdots,P_{k}\}
    =\text{In}_{\mathcal G}(R)$. This means that there is a
    $\mu_R: \mathbb F^{k}\rightarrow \mathbb F$
    such that $f_{R}=\mu_R(f_{P_{1}},\cdots,f_{P_{k}})$.
\end{itemize}
\end{defn}
Here, $f_R$ is called the {\em global encoding function} of $R$
and $\mu_R$ is called the {\em local encoding function} of
$R$.\footnote{The encoding function of a sink region in fact acts
as a ``decoding function'' since it helps to recover the source
message. We use the term ``encoding function'' because a sink
region may contains other links besides sink links.} The code
$\tilde{C}$ is called a {\em linear code} if the local encoding
functions are all linear functions.

The region graph $\mathcal G=(D,\mathcal E)$ is said to be {\em
feasible} if it has a code over some finite field. The region
decomposition $D$ is said to be {\em feasible} if $\text{RG}(D)$
is feasible. Since $\mathcal G$ is a subgraph of $\text{RG}(D)$,
if $\tilde{C}$ is a code of $\mathcal G$, then $\tilde{C}$ is a
code of $\text{RG}(D)$. Thus, if $\mathcal G$ is feasible, then
$\text{RG}(D)$ is feasible, hence $D$ is feasible.

\begin{rem}\label{RFea}
As the linear solution of $G$, a linear code of $\mathcal
G=(D,\mathcal E)$ can be described as a collection of vectors
$\tilde{C}=\{d_{R}\in\mathbb F^{2}; R\in D\}$ such that the
following two conditions hold:
\begin{itemize}
 \item[($1'$)] If $R$ is an $X_{i}$ source region or an $X_{i}$ sink region,
 then $d_{R}=\alpha_{i}~(i\in\{1,2\})$, where $\alpha_1=(1,0)$ and
 $\alpha_2=(0,1)$;
 \item[($2'$)] If $R$ is a non-source region, then $d_{R}$ is an $\mathbb F$-linear
 combination of $\{d_{P}:P\in\text{In}_{\mathcal G}(R)\}$.
\end{itemize}
\end{rem}
In this case, $d_R$ is called the \emph{(global) coding vector} of
$R$.

\begin{rem}\label{CRSG}
Clearly, a $($linear$)$ code of $\text{RG}(D^*)$ is exactly a
$($linear$)$ solution of $G$, where $D^*$ is the trivial region
decomposition of $G$. Thus, $G$ is solvable if and only if
$\text{RG}(D^*)$ is feasible.
\end{rem}

The following lemma shows that the code of a region graph can be
extended to a network coding solution of $G$.

\begin{lem}\label{FES}
Suppose $\mathcal G=(D,\mathcal E)$ is a region graph of $G$ and
$\tilde{C}=\{f_R; R\in D\}$ is a $($linear$)$ code of $\mathcal G$.
Let $C=\{f_{e}; e\in E\}$ such that $f_e=f_R$ for any $R\in D$ and
$e\in R$. Then $C$ is a $($linear$)$ solution of $G$. Moreover,
$e\in E$ is an encoding link of $C$ only if it is the leader of some
non-source region.
\end{lem}
\begin{proof}
Since $D$ is a region decomposition of $G$, for any $e\in E$, by
Definition \ref{Reg}, there is a unique $R\in D$ such that $e\in
R$. Thus, $C$ is well defined. If $e$ is the $X_i$ source link
$(i=1,2)$ and $e\in R$, then $R$ is the $X_i$ source region and by
the construction of $C$, $f_e=f_R=\alpha_i$.

Now, suppose $e\in E$ is a non-source link and
$\text{In}(e)=\{p_1,\cdots,p_k\}$. We need to prove that $f_e$ can
be computed from $f_{p_{1}},\cdots,f_{p_{k}}$. Without loss of
generality, assume $e\in R$. Then we have the following two cases.

Case 1: $e\neq\text{lead}(R)$. Then by Definition \ref{Reg}, $R$
contains an incoming link, say $p_j$, of $e$. By the construction
of $C$, we have $f_{e}=f_{p_j}$.

Case 2: $e=\text{lead}(R)$. Then $R$ is a non-source region.
Suppose $\text{In}_{\mathcal G}(R)=\{P_1,\cdots,P_\ell\}$. Then by
Definition \ref{RegG}, each $P_i$ contains an incoming link, say
$p_{i_j}$, of $e$. By the construction of $C$, $f_{e}=f_{R}$ and
$f_{p_{i_j}}=f_{P_i}~(i=1,\cdots,\ell)$. Since $f_{R}$ can be
computed from $f_{P_{1}},\cdots,f_{P_{k}}$ $($Definition
\ref{Fea}$)$, so $f_{e}$ can be computed from
$f_{p_{i_1}},\cdots,f_{p_{i_\ell}}$.

In both cases, $f_e$ can be computed from
$f_{p_{1}},\cdots,f_{p_{k}}$. By Definition \ref{LSo}, $C$ is a
solution of $G$. Checking the above discussion, if $\tilde{C}$ is
a linear code of $\mathcal G$, then $f_e$ is a linear function of
$f_{p_{1}},\cdots,f_{p_{k}}$. So $C$ is a linear solution of $G$.

During the construction of $C$, we assign the same encoding
function to all links in the same region. Thus $e\in E$ is an
encoding link of $C$ only if it is the leader of a non-source
region.
\end{proof}

\begin{figure}[htbp]
\begin{center}
\includegraphics[width=7.4cm]{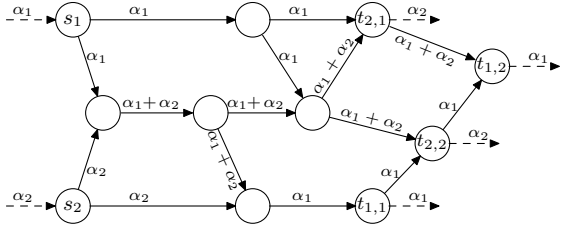}
\end{center}
\caption{A network code solution of $G_1$ obtained by extending
the code $\tilde{C}$ of $\text{RG}(D_1)$ in Example
\ref{ex-code-ext}. We depict the global encoding vector and omit
the label of each link.} \label{exm-ext-code}
\end{figure}

\begin{exam}\label{ex-code-ext}
Consider again the network $G_1$ and the region decomposition
$D_1~($See Fig. 1(a) and Fig. 2(a) respectively$)$. Let
$d_{R_1}=d_{R_4}=d_{R_5}=d_{R_8}=\alpha_1$,
$d_{R_2}=d_{R_6}=d_{R_7}=\alpha_2$, and
$d_{R_3}=\alpha_1+\alpha_2$, where $\alpha_1=(1,0)$ and
$\alpha_2=(0,1)$. By Definition \ref{Fea}, $\tilde{C}=\{d_R;R\in
D_1\}$ is a linear code of $\text{RG}(D_1)$. By Lemma \ref{FES},
$\tilde{C}$ can be extended to a linear network code solution $C$
of $G$, which is demonstrated in Fig. \ref{exm-ext-code}.
\end{exam}
\renewcommand\figurename{Fig}


The following Lemma shows that some kinds of the region
contractions can maintain the feasibility of the region graph.

\begin{lem}\label{IVC}
Suppose $\mathcal G=(D,\mathcal E)$ is a region graph of $G$ and
$\tilde{C}=\{f_{R}; R\in D\}$ is a code of $\mathcal G$. Suppose
$\{P,Q\}\subseteq D$ such that $P$ is a parent of $Q$ in $\mathcal
G$ and $f_Q$ depends only on $f_P$ $($i.e., $f_{Q}=\mu(f_{P})$ for
some $\mu:\mathbb F\rightarrow \mathbb F)$. Then $\mathcal G'$ is
feasible, where $\mathcal G'$ is the contraction of $\mathcal G$
by combining $P$ and $Q$.
\end{lem}
\begin{proof}
By Definition \ref{ConRG}, $\mathcal G'$ is a region graph of $G$
belonging to $D'$, where $D'=D\cup\{P'\}\setminus\{P,Q\}$ and
$P'=P\cup Q$. We can have the following two cases.

Case 1: $Q$ is not any sink region. Let $f_{P'}=f_P$ and
$\tilde{C'}=\{f_{R}; R\in D'\}$. It is easy to check that
$\tilde{C'}$ is a code of $\mathcal G'$. Hence, $\mathcal G'$ is
feasible.

Case 2: $Q$ is an $X_i$ sink region for some $i\in\{1,2\}$. By
Definition \ref{Fea}, $f_{Q}=\mu(f_{P})=X_i$ is a surjective
function. So $\mu$ is a surjective function. Since $\mathbb F$ is
finite, $\mu$ is bijective. We have
$f_P=\mu^{-1}(f_Q)=\mu^{-1}(X_i)$. Let $f_{P'}=X_i$ and
$\tilde{C'}=\{f_{R}; R\in D'\}$. It is easy to check that
$\tilde{C'}$ is a code of $\mathcal G'$. Hence, $\mathcal G'$ is
feasible.
\end{proof}

\begin{figure}[htbp]
\begin{center}
\includegraphics[width=7.2cm]{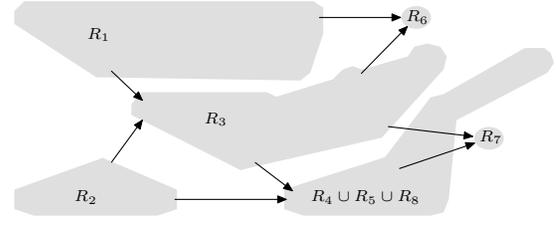}
\end{center}
\caption{The region graph $\text{RG}(D_3)$ of $G_1$, where
$D_3=\{R_1,R_2,R_3,R_4\cup R_5\cup R_8,R_6,R_7\}$, $G_1$ and
$D_1=\{R_1,\cdots,R_8\}$ are shown in Fig. 1(a) and Fig. 2(a)
respectively.} \label{ICFR}
\end{figure}

\begin{exam}\label{add-e-1}
Reconsider the network $G_1$ and the region decomposition
$D_1~($See Fig. 1(a) and Fig. 2(a) respectively$)$. We have shown
a linear code $\tilde{C}=\{d_R;R\in D_1\}$ of $\text{RG}(D_1)$ in
Example \ref{ex-code-ext}. Note that $R_4$ is a parent of $R_5$ in
$\text{RG}(D_1)$ and $f_{R_4}=f_{R_5}=\alpha_1$. Let $d_{R_4\cup
R_5}=\alpha_1$. Then we have a linear code of $\text{RG}(D_2)$,
where $D_2=\{R_1,R_2,R_3,R_4\cup R_5,R_6,R_7,R_8\}$. So $D_2$ is
feasible. Moreover, note that $R_4\cup R_5$ is a parent of $R_8$
in $\text{RG}(D_2)$ and $d_{R_4\cup R_5}=d_{R_8}=\alpha_1$. Let
$D_3=\{R_1,R_2,R_3,R_4\cup R_5\cup R_8,R_6,R_7\}$. Then $D_3$ is
also feasible. Let $d_{R_1}=d_{R_4\cup R_5\cup R_8}=\alpha_1$,
$d_{R_2}=d_{R_6}=d_{R_7}=\alpha_2$, and
$d_{R_3}=\alpha_1+\alpha_2$, where $\alpha_1=(1,0)$ and
$\alpha_2=(0,1)$. Then by Definition \ref{Fea},
$\tilde{C}=\{d_R;R\in D_1\}$ is a linear code of $\text{RG}(D_3)$.
The region graph $\text{RG}(D_3)$ is shown in Fig. \ref{ICFR}.
\end{exam}
\renewcommand\figurename{Fig}

\begin{cor}\label{UPC}
Suppose $\mathcal G=(D,\mathcal E)$ is a region graph of $G$ and
$\{P,Q\}\subseteq D$ such that $\text{In}_{\mathcal G}(Q)=\{P\}$.
Suppose $\mathcal G'$ is the contraction of $\mathcal G$ by
combining $P$ and $Q$. Then $\mathcal G$ is feasible if and only
if $\mathcal G'$ is feasible.
\end{cor}
\begin{proof}
Suppose $\mathcal G$ is feasible and $\tilde{C}=\{f_{R}; R\in D\}$
is a code of $\mathcal G$. Note that $\text{In}_{\mathcal
G}(Q)=\{P\}$. By Definition \ref{Fea}, $f_Q=\mu_Q(f_P)$, where
$\mu_Q$ is the local encoding function of $Q$. Then by Lemma
\ref{IVC}, $\mathcal G'$ is feasible.

Conversely, suppose $\mathcal G'$ is feasible and
$\tilde{C'}=\{f_{R}; R\in D'\}$ is a code of $\mathcal G'$, where
$D'=D\cup\{P\cup Q\}\setminus\{P,Q\}$ is the contraction of $D$ by
combining $P$ and $Q$. Let $f_P=f_Q=f_{P\cup Q}$. Then it is easy
to see that $\tilde{C}=\{f_{R}; R\in D\}$  is a code of $\mathcal
G$, hence $\mathcal G$ is feasible.
\end{proof}

For any region graph $\mathcal G=(D,\mathcal E)$, if there is a
non-source region, say $Q$, such that $Q$ has only one parent, say
$P$, in $\mathcal G$, then we can combine $P$ and $Q$ and obtain a
contraction $\mathcal G'$ of $\mathcal G$. This operation can be
done continuously until each non-source region has at least two
parents in $\mathcal G'$. By Corollary \ref{UPC}, $\mathcal G$ is
feasible if and only if $\mathcal G'$ is feasible. This process
can be realized by the Algorithm $1$ below, which assume that
$D=\{R_1,\cdots,R_{|D|}\}$ such that $R_i$ is the $X_i$ source
region $(i=1,2)$ and $j<\ell$ if $R_j$ is a parent of $R_\ell$.
Clearly $|D|\leq |E|$. So the runtime of Algorithm $1$ is
$O(|E|)$.
\begin{figure}[htbp]
\begin{center}
\includegraphics[width=8.0cm]{redefig5.1}
\end{center}
\end{figure}

\subsection{Feasibility and Region Labelling}
Given a region graph, the most important problem is to determine its
feasibility. Unlike the subtree graph in \cite{Fragouli06}, we
cannot use the multicast property to determine the feasibility of
region graph for 2-URMS networks, since the max-flow/min-cut
condition is not sufficient to characterize the solvability of
non-multicast networks $($e.g., see \cite{Koetter03}$)$. In this
subsection, we define a labelling operation on region graph, and
give a simple characterization on feasibility for region graph of
2-URMS networks based on such labelling operation.

\begin{defn}[Labelling On Region Graph]\label{ReL}
Let $D$ be a region decomposition of $G$ and $\mathcal G$ be a
region graph of $G$ belonging to $D$. For $i\in\{1, 2\}$, the
$X_i$ labelling operation on $\mathcal G$ is defined recursively
as follows.
\begin{itemize}
    \item[(1)] If $R$ is an $X_{i}$ source region or an $X_{i}$
    sink region, then $R$ is labelled with $X_i$;
    \item[(2)] If the parents of $R$ in $\mathcal G$ are all labelled with $X_i$,
    then $R$ is labelled with $X_i$.
\end{itemize}
\end{defn}

The $X_i$ labelling operation is well defined because $\mathcal G$
is acyclic. A region $R$ is called an {\em $X_i$ region} of
$\mathcal G$ if $R$ is labelled with $X_i$. A region which is
neither $X_1$ region nor $X_2$ region is called a \emph{coding
region} of $\mathcal G$. A region which is both $X_1$ region and
$X_2$ region is called a \emph{singular region} of $\mathcal G$.

According to Definition \ref{ReL}, the $X_i$ labelling operation
can be realized by Algorithm $2$ below, which assume that
$D=\{R_1,\cdots,R_{|D|}\}$ such that $R_i$ is the $X_i$ source
region $(i=1,2)$ and $j<\ell$ if $R_j$ is a parent of $R_\ell$.
Clearly $|D|\leq |E|$. So the runtime of Algorithm $2$ is
$O(|E|)$.
\begin{figure}[htbp]
\begin{center}
\includegraphics[width=8.0cm]{redefig5.2}
\end{center}
\end{figure}

\begin{figure}[htbp]
\begin{center}
\includegraphics[width=7.2cm]{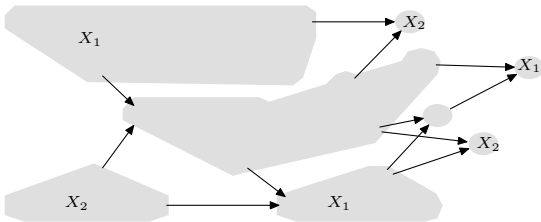}
\end{center}
\caption{Labelling operation on the region graph $\text{RG}(D_1)$
of $G_1$, where $G_1$ and $D_1$ are shown in Fig. 1(a) and Fig.
2(a) respectively.} \label{ILR1}
\end{figure}

Consider the labelling operation on the region graph
$\text{RG}(D_1)$ of network $G_1~($refer to Fig. 2(b) and Fig.
1(a) respectively$)$. The regions $R_1,R_4$ and $R_8$ are labelled
with $X_1$ since $R_1$ is the $X_1$ source region and $R_4,R_{8}$
are the $X_1$ sink region. Similarly, $R_2,R_6$ and $R_{7}$ are
labelled with $X_2$. The labelled region graph $\text{RG}(D_1)$ is
depicted in Fig. \ref{ILR1}.
\renewcommand\figurename{Fig}

\begin{exam}\label{add-e-2}
We now consider the labelling operation on another region graph of
$G_1$. Let $D_4=\{R_1,R_2,R_3\cup R_6$, $R_4,R_5,R_7,R_8\}$ as
shown in Fig. \ref{ILR1-1}(a). Likewise, $R_1,R_4,R_8$ are
labelled with $X_1$ and $R_2,R_7$ are labelled with $X_2$. Note
that $R_3\cup R_6$ is an $X_2$ sink region in $D_4$. So $R_3\cup
R_6$ is labelled with $X_2$. Moreover, since the parents of $R_4$
are all labelled with $X_2$, by Definition \ref{ReL}, $R_4$ is
labelled with $X_2$. Similarly, $R_5, R_7$ and $R_8$ are all
labelled with $X_2$. The labelled region graph $\text{RG}(D_4)$ is
depicted in Fig. \ref{ILR1-1}(b). In this case, $R_4$ and $R_8$
are two singular regions.
\end{exam}

To derive our conclusion, we need the following two lemmas.

\begin{lem}\label{Umpp}
Suppose $D$ is a region decomposition of $G$ and $\mathcal
G=(D,\mathcal E)$ is a feasible region graph of $G$. If
$\tilde{C}=\{f_{R}; R\in D\}$ is a code of $\mathcal G$ and $Q\in
D$ is an $X_i$ region $(i\in\{1,2\})$, then $f_Q$ depends only on
$X_i$. That is, there is a $\lambda_Q:\mathbb F\rightarrow \mathbb
F$ such that $f_Q(X_1,X_2)=\lambda_Q(X_i)$.
\end{lem}
\begin{proof}
We prove this lemma by induction.

If $Q$ is an $X_i$ source region or $X_i$ sink region, then by
Definition \ref{Fea}, $f_{Q}=X_i$. So $f_Q$ depends only on $X_i$.

Now, suppose $Q$ is neither $X_i$ source region nor $X_i$ sink
region. By Definition \ref{ReL}, the parents of $Q$ in $\mathcal
G$ are all $X_i$ region. Suppose $\text{In}_{\mathcal
G}(Q)=\{P_1,\cdots,P_k\}$. By induction, we can assume that
$f_{P_j}$ depends only on $X_i~(j=1,\cdots,k)$. By Definition
\ref{Fea}, $f_{Q}=\mu_{Q}(f_{P_{1}},\cdots,f_{P_{k}})$, hence
depends only on $X_i$, where $\mu_Q$ is the local encoding
function of $Q$.
\end{proof}

\begin{lem}\label{FRGC}
Suppose $D$ is a region decomposition of $G$ and $\mathcal
G=(D,\mathcal E)$ is a region graph of $G$ such that $\mathcal G$
has no singular region and each non-source region has at least two
parents in $\mathcal G$. Suppose $\tilde{C}=\{d_{R}\in \mathbb
F^2; R\in D\}$ is a collection of vectors such that
\begin{itemize}
    \item[(1)] If $R$ is an $X_i$ region, then
    $d_R=\alpha_i~(i\in\{1,2\})$, where $\alpha_1=(1,0)$ and $\alpha_2=(0,1)$;
    \item[(2)] If $\{R,Q\}\subseteq D$ such that $R,Q$ have a common child
    and are neither both $X_1$ region nor both $X_2$ region,
    then $d_{R}$ and $d_{Q}$ are linearly independent.
\end{itemize}
Then $\tilde{C}$ is a linear code of $\mathcal G$.
\end{lem}
\begin{proof}
Note that $\mathcal G$ has no singular region, $\tilde{C}$
satisfies (1') of Remark \ref{RFea}. Now take a non-source region
$R$, we only need to prove that $d_R$ is an $\mathbb F$-linear
combination of $\{d_{P}; P\in\text{In}_{\mathcal G}(R)\}$. If
there is an $i\in\{1,2\}$ such that the parents of $R$ are all
$X_i$ region, then by Definition \ref{ReL}, $R$ is an $X_i$
region. So by the assumption of $\tilde{C}$, $d_R=d_P=\alpha_i$
for all $P\in\text{In}_{\mathcal G}(R)$. Otherwise, since $R$ has
at least two parents, by the assumption (2) of $\tilde{C}$, we can
find two parents of $R$, say $P_{1}$ and $P_{2}$, such that
$d_{P_{1}}$ and $d_{P_{2}}$ are linearly independent. So $d_{P_1}$
and $d_{P_2}$ span $\mathbb F^2$. Hence $d_{R}$ is an $\mathbb
F$-linear combination of $d_{P_{1}}$ and $d_{P_{2}}$, which
completes the proof.
\end{proof}

Now, we can give a simple characterization of solvability of
region graph for 2-URMS networks.

\begin{thm}\label{CRe}
Suppose $D$ is a region decomposition of $G$ and $\mathcal
G=(D,\mathcal E)$ is a region graph of $G$ such that each
non-source region has at least two parents in $\mathcal G$. Then
$\mathcal G$ is feasible if and only if it has no singular region.
Moreover, if $\mathcal G$ is feasible, it has a linear code.
\end{thm}
\begin{proof}
If $\mathcal G$ is feasible, we shall prove by contradiction that
it has no singular region. Suppose $\mathcal G$ has a singular
region. Note that $\mathcal G$ is acyclic, we can find a singular
region $Q$ such that no parent of $Q$ is a singular region. We
declare that $Q$ contains either an $X_1$ link or an $X_2$ link or
both. $($If $Q$ contains neither $X_1$ link nor $X_2$ link, then
by Definition \ref{ReL}, all the parents of $Q$ will be singular
regions, which yields a contradiction.$)$ Without loss of
generality, we assume $Q$ contains an $X_1$ link. Since $\mathcal
G$ is feasible, $\mathcal G$ has a code $\tilde{C}=\{f_{R}; R\in
D\}$. By Definition \ref{Fea}, $f_Q(X_1,X_2)=X_1$. On the other
hand, since $Q$ is also an $X_2$ region, by Lemma \ref{Umpp},
$f_Q(X_1,X_2)$ depends only on $X_2$. A contradiction follows.
Thus, $\mathcal G$ has no singular region.

Conversely, if $\mathcal G$ has no singular region, we shall prove
that $\mathcal G$ is feasible by constructing a linear code of it.
Suppose $Q_{1},\cdots,Q_{J}$ are all coding regions of $\mathcal
G$ and $\mathbb F=\{0,c_1,,c_2,\cdots,c_{q-1}\}$ be a field of
size $q\geq J+1$, where $c_1=1$. Let $\tilde{C}=\{d_{R}\in\mathbb
F^2; R\in D\}$ such that
\begin{itemize}
    \item[(1)] If $R$ is an $X_i$ region, then $d_R=\alpha_{i}~(i\in\{1,2\})$,
    where $\alpha_1=(1,0)$ and $\alpha_2=(0,1)$;
    \item[(2)] $d_{Q_{j}}=\beta_{j}$, where $\beta_j=(1,c_j), j=1,\cdots,J$.
\end{itemize}
Note that $\alpha_1,\alpha_2$ and $\beta_j=(1,c_{j}),j=1,\cdots,J$
are mutually linearly independent. By Lemma \ref{FRGC},
$\tilde{C}$ is a linear code of $\mathcal G$. So $\mathcal G$ is
feasible.
\end{proof}

Assume $D=\{R_1,\cdots,R_{|D|}\}$ such that $R_i$ is the $X_i$
source region $(i=1,2)$ and $j<\ell$ if $R_j$ is a parent of
$R_\ell$. Suppose $\mathcal G=(D,\mathcal E)$ is a region graph of
$G$ belonging to $D$ such that each non-source region has at least
two parents in $\mathcal G$. If $\mathcal G$ is labelled with both
$X_1$ and $X_2~($according to Definition \ref{ReL}$)$, then by
Theorem \ref{CRe}, we can determine the feasibility of $\mathcal
G$ by Algorithm $3$ below. Since $|D|\leq |E|$, the runtime of
Algorithm $3$ is $O(|E|)$.
\begin{figure}[htbp]
\begin{center}
\includegraphics[width=8.0cm]{redefig5.3}
\end{center}
\end{figure}

Moreover, the proof of Theorem \ref{CRe} and Lemma \ref{FES} give a
construction method for linear solution of $G$, which we summarize
in Algorithm $4$. Since $|D|\leq |E|$, the runtime of Algorithm $4$
is $O(|E|)$.
\begin{figure}[htbp]
\begin{center}
\includegraphics[width=8.0cm]{redefig5.4}
\end{center}
\end{figure}

Let $\mathcal G=(D,\mathcal E)$ be a region graph of $G$ belonging
to $D$. We can determine its feasibility as follows: 1) If there is
a non-source regions $R\in D$ such that $R$ has only one parent in
$\mathcal G$, then use Algorithm $1$ to contract $\mathcal G$ to a
region graph $\mathcal G'=(D',\mathcal E')$ such that each
non-source region in $\mathcal G'$ has at least two parents; 2) Use
Algorithm $2$ to perform $X_i$ labelling on $\mathcal G', i=1,2$; 3)
Use Algorithm $3$ to determine the feasibility of $\mathcal G'$. By
Lemma \ref{UPC}, $\mathcal G$ is feasible if and only if $\mathcal
G'$ is feasible. So once we determined the feasibility of $\mathcal
G'$, we have also determined the feasibility of $\mathcal G$.

\renewcommand\figurename{Fig}
\begin{figure}[htbp]
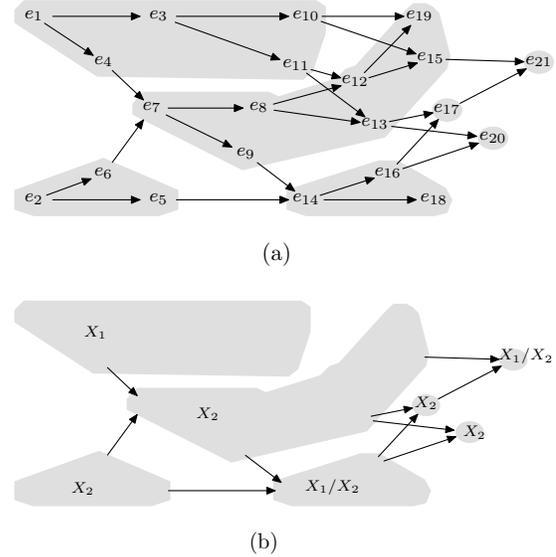

\begin{center}
\includegraphics[width=7.2cm]{redefig2.11}
\includegraphics[width=7.2cm]{redefig2.8}
\end{center}
\caption{(a) The region decomposition $D_4$ of $G_1$. (b)
Labelling operation on the region graph $\text{RG}(D_4)$, where
$G_1$ is shown in Fig. 1(a) and $D_4$ is described in Example
\ref{add-e-2}.} \label{ILR1-1}
\end{figure}

\section{Time Complexity for a Solution}
In this section, we give $O(|E|)$ time algorithms to determine
solvability and to construct network coding solutions of the
2-URMS network $G$. We shall prove that $G$ has a unique region
decomposition $D^{**}$, called the basic region decomposition,
which can be obtained in time $O(|E|)$, and $G$ is solvable if and
only if $\text{RG}(D^{**})$ is feasible. First, we give the
definition of basic region decomposition.

\begin{defn}[Basic Region Decomposition]\label{BRD}
Let $D$ be a region decomposition of $G$. $D$ is said to be a
basic region decomposition of $G$ if the following two conditions
hold.
\begin{itemize}
    \item [(1)] For any region $R\in D$ and any link $e\in
    R\setminus\{\text{lead}(R)\}$, $\text{In}(e)\subseteq R$;
    \item [(2)] Each non-source region $R$ in $D$ has at least two
    parents in $\text{RG}(D)$.
\end{itemize}
If $D$ is a basic region decomposition of $G$, then $\text{RG}(D)$
is called a basic region graph of $G$.
\end{defn}

The following two examples demonstrate this notion.

\begin{exam}\label{EBRG1}
Consider again the example network $G_1$ in Fig. 1(a). Let
$Q_{1}=\{e_{1},e_{3},e_{4},e_{10},e_{11}\}$,
$Q_{2}=\{e_{2},e_{5},e_{6}\}$, $Q_{3}=\{e_{7},e_{8},e_{9}\}$,
$Q_4=\{e_{12}\}$, $Q_5=\{e_{13}\}$,
$Q_6=\{e_{14},e_{16},e_{18}\}$, $Q_7=\{e_{15}\}$,
$Q_{8}=\{e_{17}\}$, $Q_{9}=\{e_{19}\}$, $Q_{10}=\{e_{20}\}$ and
$Q_{11}=\{e_{21}\}$. By Definition \ref{BRD}, we can easily check
that $D^{**}=\{Q_{1},\cdots,Q_{11}\}$ is a basic region
decomposition of $G_1$.
\end{exam}

\renewcommand\figurename{Fig}
\begin{figure}[htbp]
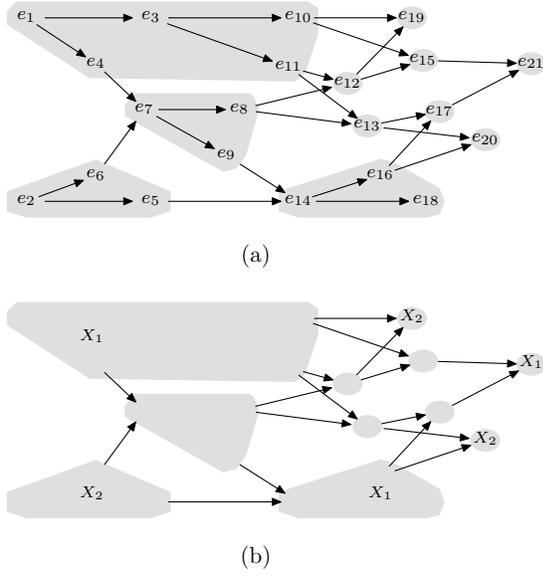

\begin{center}
\includegraphics[width=7.2cm]{redefig3.2}
\includegraphics[width=7.2cm]{redefig3.3}
\end{center}
\caption{(a) is the basic region decomposition $D^{**}$ of
$G_1~($See Fig. 1(a)$)$; (b) is the labelled region graph
$RG(D^{**})$ of $G_1$.} \label{BRG1}
\end{figure}

\begin{exam}\label{EBRG2}
We consider another example network $G_2$ in Fig. \ref{BRG2}. Let
$R_{1}=\{e_{1},e_{3},e_{4},e_{5},e_{7},e_{8},e_{9}\}$,
$R_{2}=\{e_{2},e_{6}\}$,
$R_{3}=\{e_{10},e_{11},e_{12},e_{13},e_{15}\}$, $R_4=\{e_{14}\}$,
$R_{5}=\{e_{16}\}$. It can be checked that
$D^{**}=\{R_{1},R_{2},R_{3},R_{4},R_{5}\}$ is a basic region
decomposition of $G_2$.
\end{exam}

For the example network $G_1$, note that $D_1,D_2,D_3$ and
$D_4~($see Example \ref{ERG1}, \ref{EG1}, \ref{add-e-1} and
\ref{add-e-2}$)$ are not basic region decomposition since they do
not satisfy condition (1) of Definition \ref{BRD}. The trivial
region decomposition $D^*$ is either not a basic region
decomposition since it does not satisfy condition (2) of
Definition \ref{BRD}.

In general, let $E=\{e_1,e_2,\cdots,e_{|E|}\}$ be indexed as in
Remark \ref{indexE}. We can then obtain a basic region decomposition
of $G$ by the following Algorithm $5$.
\begin{figure}[htbp]
\begin{center}
\includegraphics[width=8.0cm]{redefig5.5}
\end{center}
\end{figure}
It is easy to prove the correctness of Algorithm $5$ by Definition
\ref{BRD}. Moreover, note that Algorithm $5$ makes
$|\text{In}(e_j)|$ times comparisons for each $e_j~(j\geq 3)$. So
its runtime is $O(|E|)$.

\renewcommand\figurename{Fig}
\begin{figure}[htbp]
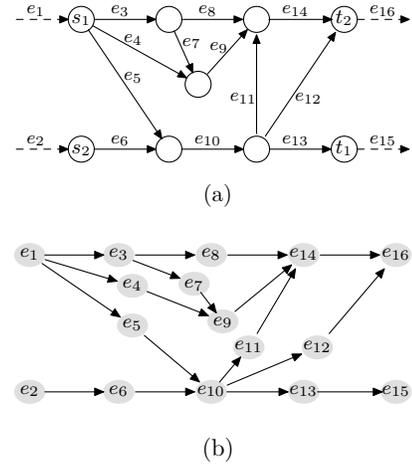

\begin{center}
\includegraphics[width=5.3cm]{redefig4.1}
\includegraphics[width=5.3cm]{redefig4.2}
\end{center}
\caption{(a) is an example network $G_2$, where the imaginary
links $e_1$ and $e_2$ are the $X_1$ source link and $X_2$ source
link respectively, and the imaginary links $e_{15}$ and $e_{16}$
are the $X_1$ sink link and $X_2$ sink link respectively. (b) is
the line graph $L(G_2)$ of $G_2$.} \label{BRG2}
\end{figure}

\renewcommand\figurename{Fig}
\begin{figure}[htbp]
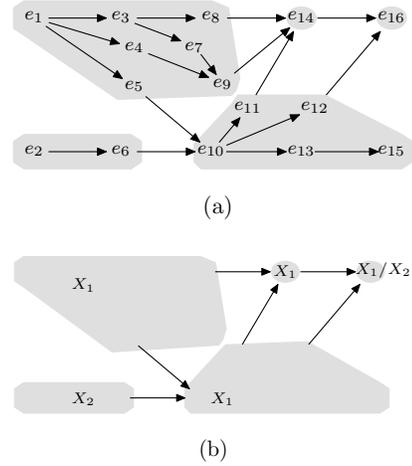

\begin{center}
\includegraphics[width=5.3cm]{redefig4.4}
\includegraphics[width=5.3cm]{redefig4.3}
\end{center}
\caption{(a) is the basic region decomposition $D^{**}$ of $G_2$;
(b) The labelled region graph $RG(D^{**})$.} \label{BRG22}
\end{figure}

An interesting result is the following theorem.

\begin{thm}\label{U-ReDe}
$G$ has a unique basic region decomposition, hence has a unique
basic region graph.
\end{thm}
\begin{proof}
Suppose $D=\{R_1,\cdots,R_K\}$ and $D'=\{Q_1,\cdots,Q_L\}$ are
both basic region decomposition of $G$. We shall prove that
$D=D'$.

First, we prove that any $R_i\in D$ is contained in some region in
$D'$. Let $E=\{e_1,e_2,\cdots,e_{|E|}\}$ be indexed as in Remark
\ref{indexE}. Assume
$R_{i}=\{e_{i_{1}},e_{i_{2}},\cdots,e_{i_{n}}\}$ such that
$i_{1}<i_{2}<\cdots<i_{n}$. Then $e_{i_{1}}=\text{lead}(R)$.
$($Otherwise, by Definition \ref{Reg}, there is an $e_{i_{j}}\in
R_i$ and $e_{i_{j}}$ is an incoming link of $e_{i_{1}}$. By Remark
\ref{indexE}, $i_j<i_1$, which contradicts to the assumption that
$i_{1}<i_{2}<\cdots<i_{n}.)$ Since $D'=\{Q_1,\cdots,Q_L\}$ is a
basic region decomposition of $G$, there is a $Q\in D'$ such that
$e_{i_1}\in Q$. Without loss of generality, assume $e_{i_{1}}\in
Q_{1}$. Then $R_{i}\subseteq Q_{1}$. $($Otherwise, there exists an
$e_{i_{k}}\in R_{i}$ such that
$\{e_{i_{1}},\cdots,e_{i_{k-1}}\}\subseteq Q_{1}$ and
$e_{i_{k}}\in Q_{j}~(j\neq 1)$. By Remark \ref{indexE} and (1) of
Definition \ref{BRD}, $\text{In}(e_{i_{k}})\subseteq
\{e_{i_{1}},\cdots,e_{i_{k-1}}\}\subseteq Q_1$. Since
$D'=\{Q_1,\cdots,Q_L\}$ is a basic region decomposition of $G$,
$Q_{1}\cap Q_{j}=\emptyset$. So $Q_j$ contains no incoming link of
$e_{i_k}$. Then by Definition \ref{Reg},
$e_{i_{k}}=\text{lead}(Q_{j})$ and $Q_{1}$ is the only parent of
$Q_{j}$, which contradicts to (2) of Definition \ref{BRD}.$)$

Similarly, we can prove that $Q_{1}\subseteq R_{\ell}$ for some
$R_{\ell}\in D$. So $R_{i}\subseteq R_{\ell}$. Since
$D=\{R_1,\cdots,R_K\}$ is a basic region decomposition of $G$, if
$R_{i}\neq R_{\ell}$, then $R_{i}\cap
R_{\ell}=\emptyset~($Definition \ref{Reg}$)$. So we have
$R_{i}=R_{\ell}=Q_{1}$. Note that $R_i$ can be arbitrarily chosen
from $D$, we have $D\subseteq D'$.

Symmetrically, we have $D'\subseteq D$.

Thus $D'=D$, which derives the uniqueness of the basic region
decomposition of $G$.
\end{proof}

In the sequel, we will always use $D^{**}$ to denote the basic
region decomposition of $G$. And the basic region graph of $G$ is
$\text{RG}(D^{**})$. Now, we can discuss the solvability of $G$.

The basic region decomposition, as well as the basic region graph,
is a new notion in our approach, which is not introduced in
\cite{Fragouli06}. The most important property of the basic region
graph is that the solvability of $G$ is equivalent to the
feasibility of the basic region graph of $G$, which is shown in the
following theorem.

\begin{thm}\label{BReg}
Let $D^{**}$ be the basic region decomposition of $G$. Then $G$ is
solvable if and only if $\text{RG}(D^{**})$ is feasible.
\end{thm}
\begin{proof}
Suppose $\text{RG}(D^{**})$ is feasible. Then $\text{RG}(D^{**})$
has a code $\tilde{C}=\{f_R; R\in D\}$. By Lemma \ref{FES},
$\tilde{C}$ can be extended to a solution of $G$. So $G$ is
solvable.

Conversely, suppose $G$ is solvable. By Remark \ref{CRSG}, the
trivial region graph $\text{RG}(D^*)$ is feasible. According to
Algorithm 5, $D^{**}$ is in fact obtained from $D^{*}$ by a series
of region contractions, i.e., if the region $R_{e_j}=\{e_j\}$ has
a unique parent $R_k$ then combine $R_{k}$ and $R_{e_j}$. Hence
its feasibility remains unchanged (Corollary \ref{UPC}). So
$\text{RG}(D^{**})$ is feasible.
\end{proof}

For the example network $G_1$, we have obtained its basic region
decomposition in Fig. \ref{BRG1}(a). Using Algorithm 2, we obtain
a labelled region graph $\text{RG}(D^{**})$ of $G_1$ as shown in
Fig. \ref{BRG1}(b). One can see that $\text{RG}(D^{**})$ has no
singular region, and hence is feasible by Theorem \ref{CRe}.
Hence, by Theorem \ref{BReg}, $G_1$ is solvable.

Consider the basic region decomposition of the network $G_2~($See
Fig. 9(a)$)$. Using Algorithm 5 and Algorithm 2, we obtain a
labelled region graph $\text{RG}(D^{**})$ of $G_2$ as shown in
Fig.\ref{BRG22}(b). One can see that $\text{RG}(D^{**})$ has a
singular region $R_5$ and hence is not feasible by Theorem
\ref{CRe}. Following Theorem \ref{BReg}, $G_2$ is not solvable.

For a given 2-URMS network $G$, the determination of the
solvability and construction of a linear solution can be realized
by the following three steps: 1) Use Algorithm 5 to obtain its
basic region decomposition $D^{**}$; 2) Use Algorithm 2 and
Algorithm 3 to determine the feasibility of the basic region graph
$\text{RG}(D^{**})$; 3) If $G$ is solvable, then use Algorithm 4
to construct a linear code on $\text{RG}(D^{**})$ and extend it to
a linear solution of $G$. We have shown that all algorithms are
with runtime $O(|E|)$, so we have the following theorem.

\begin{thm}\label{PSRG}
The solvability of $G$ can be determined in $O(|E|)$ time.
Furthermore, if $G$ is solvable, a linear solution of $G$ can be
constructed in time $O(|E|)$.
\end{thm}

Combining Theorem \ref{CRe} and Theorem \ref{BReg}, we have a
necessary and sufficient condition on the solvability of 2-URMS
networks based on region decomposition and region labelling. In
\cite{Wang110}, the authors gave another two necessary and
sufficient conditions. One is characterized by the ``critical
1-edge cuts'' and the other is characterized by ``paths with
controlled edge-overlap''. According to their results, determining
the solvability of 2-URMS networks needs searching for the
critical 1-edge cuts or edge-disjoint paths, for which the fastest
runtime is also $O(|E|)$. However, the code construction in
\cite{Wang110} is an exponential-time algorithm. In fact, the code
construction method in \cite{Wang110} consists of two stages.
First, construct a ``strengthened generic linear code multicast
(LCM)''. Then perform ``reset-to-X'' (resp. ``reset-to-Y'')
operations on the critical 1-edge cuts. This method is based on
the construction of LCMs in \cite{Li03}, which is an
exponential-time algorithm. In this paper, we decompose the
network into disjoint regions and assign the encoding vector of
each region in a decentralized manner, which is an $O(|E|)$ time
algorithm by Theorem \ref{PSRG}.

All notions about region decomposition, including region
decomposition, region graph, codes on region graph, region
contraction, region labelling, basic region decomposition and
basic region graph, are applicable to arbitrary directed acyclic
networks. If $G$ is a directed acyclic network with $k~(k>2)$
unit-rate multicast sessions, the region graph of $G$ will have
$k$ source regions. Correspondingly, there are $k$ types of sink
regions and $k$ types of region labelling. Also, we can prove that
$G$ has a unique basic region decomposition $($Similar to Theorem
4.4$)$ and $G$ is solvable if and only if its basic region graph
is feasible$($Similar to Theorem 4.5$)$. Unfortunately, we can't
give a characterization of feasibility of region graph as Theorem
\ref{CRe} when $k>2$. In fact, given a region graph $\mathcal G$,
the condition that $\mathcal G$ has no singular region is only a
necessary but not a sufficient condition of feasibility of
$\mathcal G$. The difficulty lies in that one cannot assign
decentralized code when the number of different information flows
is more than $2$. Thus, we need to give more intensive analysis on
the structure of region graph. For example, we can consider {\em
higher dimensional regions}. The region decomposition approach
used for two unicast network with rate (1,2) can be found in
\cite{Song12}. To characterize the feasibility of region graph for
more general networks is still an open problem.

\section{The Number of Encoding Links}
Throughout this section, we assume that $G=(V,E)$ is a 2-URMS
network with two disjoint sink sets $T_1$ and $T_2$, i.e., $T_1\cap
T_2=\emptyset$\footnote{If there is a $t\in T_1\cap T_2$, we can add
two additional nodes $t'$ and $t''$ with two additional links
$(t,t')$ and $(t,t'')$. Replace $t$ by $t'$ in $T_1$ and $t''$ in
$T_2$ respectively, we get a new network such that any network
coding solution for the original network can be mapped bijectively
to a network coding solution for the new one without changing the
encoding complexity.}. Since $T_1\cap T_2=\emptyset$, each sink
corresponds to a sink link, hence the number of sinks is equal to
the number of sink links. Moreover, since each sink region contains
at least one sink link, then the number of sink regions is not
greater than the number of sink links. Thus, we have the following
remark.

\begin{rem}\label{num-sink}
The number of sink regions is not greater than the number of
sinks.
\end{rem}

In this section, we shall prove the following theorem.

\begin{thm}\label{encn} Let $G$ be a solvable 2-URMS
network with $N$ sinks, then $G$ has a network coding solution
with at most max$\{3,2N-2\}$ encoding links. There exist instances
which achieve this bound.
\end{thm}

Before we proceed with the proof of Theorem \ref{encn}, we introduce
the concept of {\em minimal feasible region graph}, which is
analogue to the minimal subtree graph in \cite{Fragouli06}.

\begin{defn}[Minimal Feasible Region Decomposition]\label{MFR}
Let $D$ be a feasible region decomposition of $G$. We say that $D$
is a minimal feasible region decomposition of $G$ if combining any
adjacent regions in $\text{RG}(D)$ results in a contraction of $D$
which is not feasible.
\end{defn}

\begin{defn}[Minimal Feasible Region Graph]\label{MFea}
Let $D$ be a region decomposition of $G$ and $\mathcal G=(D,\mathcal
E)$ be a feasible region graph of $G$ belonging to $D$. We say that
$\mathcal G$ is a minimal feasible region graph of $G$ if the
following two conditions hold:
\begin{itemize}
    \item [(1)] $D$ is a minimal feasible region decomposition of
    $G$;
    \item [(2)] Deleting any edge of $\mathcal G$ results in a subgraph
    of $\mathcal G$ which is not feasible.
\end{itemize}
\end{defn}

Clearly, the condition (1) in Definition \ref{MFea} is equivalent
to the condition: (1') Combining any adjacent regions in $\mathcal
G$ results in a contraction of $\mathcal G$ which is not feasible.

For any solvable network $G$, by Theorem \ref{BReg}, the basic
region graph $\text{RG}(D^{**})$ of $G$ is feasible. If
$\text{RG}(D^{**})$ is not minimal feasible, one can always get a
smaller feasible region graph, i.e., with less links and/or less
nodes by deleting edges and/or combining nodes of
$\text{RG}(D^{**})$. Once we cannot further perform
deleting/combining process, we get a minimal feasible region graph
of $G$. The solution derived from the minimal feasible region
graph will have less (or equal) encoding links than the solution
derived from the original region graph. We call the solution
constructed from a minimal feasible region graph as an {\em
optimal solution} of $G$.

\begin{exam}
Consider the feasible region graphs $\text{RG}(D_1)$,
$\text{RG}(D_2)$ and $\text{RG}(D_3)$ of the example network
$G_1($refer to Fig. 2, Fig. 3 and Fig. 5 respectively$)$. It can
be checked that both $\text{RG}(D_1)$ and $\text{RG}(D_2)$ are not
minimal feasible, while $\text{RG}(D_3)$ is minimal feasible. The
solution of $G_1$ given in Example \ref{ex-code-ext} can also be
viewed as being extended from the code of $\text{RG}(D_3)$ given
in Example \ref{add-e-1}, which is an optimal solution of $G_1$
and has only $4$ encoding links, i.e., the leader of $R_3$,
$R_4\cup R_5\cup R_8$, $R_6$ and $R_7$.
\end{exam}

\begin{rem}\label{nu-mrd}
In general, the minimal feasible region decomposition is not
unique. For example, for the network $G_1$, let
$P_1=\{e_{1},e_{3},e_{4},e_{10},e_{11},e_{15},e_{21}\},
P_2=\{e_{2},e_{5},e_{6}\},
P_3=\{e_{7},e_{8},e_{9},e_{12},e_{13},e_{17}\},
P_4=\{e_{14},e_{16},e_{18}\}, P_5=\{e_{19}\}$, $P_6=\{e_{20}\}$.
By Definition \ref{MFR}, we can easily check that
$D_5=\{P_1,P_2,P_3,P_4,P_5,P_6\}$ is a minimal feasible region
decomposition of $G_1$. Thus, $D_3$ and $D_5$ are two different
minimal feasible region decompositions of $G_1$.
\end{rem}

We assume that $D=\{R_1,\cdots,R_{|D|}\}$ such that $R_i$ is the
$X_i$ source region $(i=1,2)$ and $j<\ell$ if $R_j$ is a parent of
$R_\ell$. The process of reducing a feasible region graph to a
minimal feasible region graph can be summarized in Algorithm $6$.
\begin{figure}[htbp]
\begin{center}
\includegraphics[width=8.0cm]{redefig5.6}
\end{center}
\end{figure}

For each non-source region $R_j\in D$, Algorithm $6$ makes at most
$|\text{In}(\text{lead}(R_j)|$ times verifications of (1) and (2)
of Definition \ref{MFea}. Each time of the verification can be
done by Algorithm $2$ and Algorithm $3$ with $O(|E|)$ time. Note
that $|\text{In}(R_j)|\leq|\text{In}(\text{lead}(R_j))|$, so
Algorithm $5$ is also a polynomial time algorithm. Once we obtain
the minimal feasible region graph by Algorithm $6$, an optimal
solution of $G$ can be constructed by Algorithm $4$ in $O(|E|)$
time. Thus, construction of an optimal solution of $G$ can be
completed in $O(|E|)$ time.

Similar to the minimal subtree graphs in \cite{Fragouli06}, the
minimal feasible region graphs also have some interesting
properties.

\begin{thm}\label{Min}
Let $\mathcal G=(D,\mathcal E)$ be a minimal feasible region graph
of $G$. The following statements hold:
\begin{itemize}
    \item [1)] Any non-source region has exactly two parents.
    \item [2)] Two regions which are adjacent or have a common child
    cannot be both $X_1$ region nor both $X_2$ region.
    \item [3)] Two adjacent coding regions have a common child.
    \item [4)] If a coding region $R$ is adjacent to an $X_i$ region
    $(i\in\{1,2\})$, then there exists an $X_i$ region, say $Q$, such
    that $R$ and $Q$ have a common child.
\end{itemize}
\end{thm}
\begin{proof}
1) Suppose $Q$ is a non-source region of $G$. We need to exclude the
case that $Q$ has no parent and the case that $Q$ has more than two
parents in $\mathcal G$. If $Q$ has no parent in $\mathcal G$, then
$Q$ has no impact on the feasibility of $\mathcal G$. So for any
parent $P$ of $Q$ in $\text{RG}(D)$, we can combine $P$ and $Q$ and
obtain a contraction $D'$ of $D$ such that $D'$ is feasible, which
contradicts to the minimality of $\mathcal G$. Now, suppose $Q$ has
more than two parents. Let $\tilde{C}=\{d_{R}\in \mathbb F^2; R\in
D\}$ be a code of $\mathcal G$ as in the proof of Theorem \ref{CRe}.
Since $\mathbb F^2$ is of dimension two, there must be two parents
of $Q$, say $P_1$ and $P_2$, such that $d_Q$ is an $\mathbb
F$-linear combination of $d_{P_{1}}$ and $d_{P_{2}}$. Then delete
the edge(s) between $Q$ and all its other parents, we obtain a
feasible subgraph with code $\tilde{C}$, which contradicts to the
minimality of $\mathcal G$. So $Q$ has exactly two parents, and 1)
holds.

2) Suppose $P$ and $Q$ are both $X_1$ regions (or both $X_2$
regions) and $\tilde{C}$ is a code of $\mathcal G$ as in the proof
of Theorem \ref{CRe}. Then $d_P=d_Q=\alpha_1~($resp.
$d_P=d_Q=\alpha_2)$. If $P$ and $Q$ are adjacent, by Lemma
\ref{IVC}, $D$ can be contracted by combining $P$ and $Q$ without
changing its feasibility. Similarly, if $P$ and $Q$ have a common
child $R$, then by deleting the edge between $Q$ and $R$ we obtain
a subgraph $\mathcal G'$ of $\mathcal G$ such that $\tilde{C}$ is
still a code of $\mathcal G'$. In both cases we derive a
contradiction to the minimality of $\mathcal G$, hence 2) holds.

3) Suppose $P, Q\in D$ are two adjacent coding regions which have
no common child. Let $\tilde{C}$ be the code of $\mathcal G$ as in
the proof of Theorem \ref{CRe}. We alter $d_Q$ by letting
$d_Q=d_P$, and keep the rest of global coding vectors unchanged.
Since $P$ and $Q$ have no common child, this assignment is still a
code of $\mathcal G~($Lemma \ref{FRGC}$)$. By Lemma \ref{IVC}, $D$
can be contracted by combining $P$ and $Q$ without changing the
feasibility, which contradicts to the minimality of $\mathcal G$,
hence 3) holds.

4) Suppose $R$ is adjacent to an $X_i$ region $P$ and $R$ has no
common child with any $X_i$ region. Let $\tilde{C}$ be the code of
$\mathcal G$ as in the proof of Theorem \ref{CRe}. We alter $d_R$
by letting $d_R=\alpha_i$, and keep the rest of global encoding
kernels unchanged. Since $R$ has no common child with any $X_i$
region, this assignment is still a code of $\mathcal G~($Lemma
\ref{FRGC}$)$. By Lemma \ref{IVC}, $D$ can be contracted by
combining $R$ and $P$ without changing the feasibility, which
contradicts to the minimality of $\mathcal G$, hence 4) holds.
\end{proof}

For the sake of convenience, we say that a region $Q$ is an
$X_i$-\emph{parent} (or an $X_i$-\emph{child}) of a region $R$ if
$Q$ is an $X_i$ region and a parent $($resp. a child$)$ of $R$.
The following corollary further shows some interesting properties
of the minimal feasible region graphs.

\begin{cor}\label{Min2}
Let $\mathcal G=(D,\mathcal E)$ be a minimal feasible region graph
of $G$. The following items hold.
\begin{itemize}
    \item [1)] An $X_i$ region is either an $X_{i}$ source region
    or an $X_{i}$ sink region $(i\in\{1, 2\})$. 
    \item [2)] A coding region has at least two children which are sink regions.
    \item [3)] If $Q\in D$ is a coding region such that no child of $Q$
    is a coding region, then $Q$ has two children, say $Q_{1}$ and $Q_{2}$,
    such that $Q_{i}$ is an $X_{i}$ sink region and $Q_{i}$ has an
    $X_{j}$-parent $(i\in\{1,2\}$ and $j\in\{1,2\}\setminus\{i\})$.
\end{itemize}
\end{cor}
\begin{proof}
1) Let $R\in D$ be an $X_i$ region. If $R$ is neither an $X_{i}$
source region nor an $X_{i}$ sink region, i.e. $R$ contains
neither $X_{i}$ source link nor $X_{i}$ sink link, then the
parents of $R$ are all $X_i$ region (Definition \ref{ReL}), which
contradicts to 2) of Theorem \ref{Min}.

2) Let $R$ be a coding region. Then by 1) of Theorem \ref{Min}, $R$
has two parents, say $P_{1}$ and $P_{2}$. By 2) of Theorem
\ref{Min}, they are neither both $X_1$ region nor both $X_2$ region.
We divide the discussion into the following three cases.

Case 1: $P_{1}$ is a coding region and $P_{2}$ is an $X_i$ region
$(i\in\{1,2\})$. First, consider $P_1$ and $R$. By 3) of Theorem
\ref{Min}, $P_1$ and $R$ have a common child $Q_1$. If $Q_1$ is an
$X_{j}$ region for some $j\in\{1, 2\}$, we halt. Else, if $Q_1$ is a
coding region, then by 3) of Theorem \ref{Min}, $R$ and $Q_1$ have a
common child $Q_2$. Similarly, either $Q_2$ is an $X_j$ region for
some $j\in\{1, 2\}$, or $R$ and $Q_2$ have a common child $Q_3$.
Since $\mathcal G$ is a finite graph, we can finally find an
$X_j$-child $Q_m$ of $R$. By claim 1), $Q_m$ is a sink region.

Next, consider $P_2$ and $R$. Without loss of generality, we
assume that $P_{2}$ is an $X_1$ region. By 4) of Theorem
\ref{Min}, there exists an $X_1$ region $P$ such that $R$ and $P$
have a common child $W_1$. If $W_1$ is an $X_{j}$ region for some
$j\in\{1, 2\}$, we halt. Else, if $W_1$ is a coding region, then
by  3) of Theorem \ref{Min}, $R$ and $W_1$ have a common child
$W_2$. Similarly, either $W_2$ is an $X_j$ region for some
$j\in\{1, 2\}$ or $R$ and $W_2$ have a common child $W_3$. Since
$\mathcal G$ is a finite graph, we can finally find an $X_j$-child
$W_n$ of $R$. By claim 1), $W_n$ is a sink region.

Note that $P_{1}$ is a coding region and $P$ is an $X_1$ region.
So $P_1\neq P$ and $\text{In}_{\mathcal
G}(W_1)\neq\text{In}_{\mathcal G}(Q_1)$. Thus, $Q_1\neq W_1$. By
comparing their parents, we have $Q_k\neq W_\ell,
k\in\{1,\cdots,m\}$ and $\ell\in\{1,\cdots,n\}$. So $Q_m\neq W_n$.
Thus, $Q_m$ and $W_n$ are two children of $R$ which are both sink
region.

Case 2: Both $P_{1}$ and $P_{2}$ are coding regions. Similar to
case 1, we can find two children of $R$ which are both sink
region.

Case 3: $P_{1}$ is an $X_1$ region and $P_{2}$ is an $X_2$ region.
Similar to case 1, we can find two children of $R$ which are both
sink region.

In all cases, we can find two children of $R$ which are both sink
region.

3) By claim 2), $Q$ has an $X_i$-child $Q_0$ for some
$i\in\{1,2\}$. Without loss of generality, we assume that $Q_0$ is
an $X_2$ region. By 4) of Theorem \ref{Min}, there is a region
$Q_1$ which is a common child of $Q$ and an $X_2$ region. By 2) of
Theorem \ref{Min}, $Q_1$ is not an $X_2$ region. Since no child of
$Q$ is a coding region, so $Q_1$ is an $X_1$ region.

Now consider $Q$ and $Q_1$. By 4) of Theorem \ref{Min}, there is a
region $Q_2$ which is a common child of $Q$ and an $X_1$ region.
By 2) of Theorem \ref{Min}, $Q_2$ is not an $X_1$ region. Since no
child of $Q$ is a coding region, so $Q_2$ is an $X_2$ region. By
1), $Q_1$ and $Q_2$ are both sink regions, which completes the
proof.
\end{proof}

\renewcommand\figurename{Fig}
\begin{figure}[htbp]
\begin{center}
\includegraphics[height=8cm]{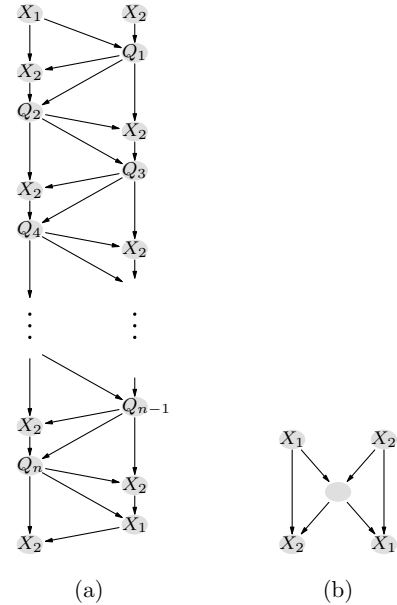}
\end{center}
\caption{Region graphs achieving the bound in Theorem \ref{Min3}:
(a) is a region graph with $N\geq 3$ sink regions and $n=N-2$
coding regions $\{Q_1,Q_2,\cdots,Q_n\}$; (b) is a region graph
with $N=2$ sink regions and $n=1$ coding regions.} \label{rgab}
\end{figure}

\begin{thm}\label{Min3}
Let $\mathcal G=(D,\mathcal E)$ be a minimal feasible region graph
of $G$ with $n$ coding regions. Then $n\leq\max\{1,N-2\}$.
\end{thm}
\begin{proof}
We define the C-S edge of $\mathcal G$ as the edge which starts at
a coding region and terminates at a sink region. Let $J$ be the
number of C-S edges and $K$ be the number of sink regions of $D$.

Suppose $D$ has $n\geq 2$ coding regions, we shall prove $n\leq
N-2$ by counting $J$ in two different ways. Firstly, note that
$\mathcal G$ is acyclic, its vertices can be ordered in an
upstream-to-downstream fashion, i.e., $\forall\{R,R'\}\subseteq
D$, $R<R'$ if $R$ is a parent of $R'$. Let $P$ and $Q$ be the
greatest two coding regions, i.e., for any coding region
$R\notin\{P,Q\}$, $R<P$ and $R<Q$. Without loss of generality,
assume $P<Q$. Then no child of $Q$ is a coding region. By 3) of
Corollary \ref{Min2}, $Q$ has two children $Q_1$ and $Q_2$, such
that $Q_i$ is an $X_i$ sink region and has an $X_j$-parent
$U_i~(i\in\{1,2\}$ and $j\in\{1,2\}\setminus\{i\})$. We
distinguish the following two cases to discuss.

Case 1: $Q$ is a child of $P$.

By 1) of Theorem \ref{Min}, each sink region has exactly two
parents. So there are exactly $2K$ edges terminate at a sink
region. Among these $2K$ edges, $(U_1,Q_1)$ and $(U_2,Q_2)$ are
not C-S edges. So there are at most $2K-2$ C-S edges. Thus,
\begin{equation} J\leq 2K-2.\tag{1} \end{equation}

On the other hand, by 2) of Corollary \ref{Min2}, each coding
region has at least two children which are sink region. So except
$Q$, the $n-1$ coding regions supply at least $2(n-1)$ C-S edges.
Now consider the coding region $Q$. Note that $P$ and $Q$ are
adjacent, by 3) of Theorem \ref{Min}, $P$ and $Q$ have a common
child, say $Q_3$. Since $P$ and $Q$ are the greatest two coding
regions, $Q_3$ could not be a coding region. By 1) of Corollary
\ref{Min2}, $Q_3$ is a sink region. By comparing the parent set,
we have $Q_3\notin \{Q_1,Q_2\}$. So $(Q,Q_1),(Q,Q_2)$ and
$(Q,Q_3)$ are three C-S edges. Hence, there are at least
$2(n-1)+3$ C-S edges. We have
\begin{equation} 2(n-1)+3\leq J.\tag{2} \end{equation}

Combining equations (1) and (2), we have $n\leq K-3/2$. Note that
$n$ is an integer. Then $n\leq K-2$. By Remark \ref{num-sink},
$K\leq N$. So $n\leq N-2$.

Case 2: $P$ and $Q$ are not adjacent.

Then no child of $P$ is a coding region. Similar to $Q$, we can
prove that $P$ has two children $P_1$ and $P_2$, such that $P_i$
is an $X_i$ sink region and has an $X_j$-parent $W_i~(i\in\{1,2\}$
and $j\in\{1,2\}\setminus\{i\})$. By 1) of Theorem \ref{Min}, each
sink region has exactly two parents. So there are exactly $2K$
edges terminate at a sink region. Among these $2K$ edges,
$(U_1,Q_1),(U_2,Q_2),(W_1,P_1)$ and $(W_2,P_2)$ are not C-S edges.
So there are at most $2K-4$ C-S edges. Thus,
\begin{equation} J\leq 2K-4.\tag{3} \end{equation}

On the other hand, by 2) of Corollary \ref{Min2}, $n$ coding
regions supply at least $2n$ C-S edges. So \begin{equation} 2n\leq
J.\tag{4} \end{equation}

Combining equations (3) and (4), we have $n\leq K-2$. By Remark
\ref{num-sink}, $K\leq N$. So $n\leq N-2$.

Thus, we proved that $n\leq N-2$ if $n\geq 2$. That is, $n\leq 1$
or $n\leq N-2$, i.e., $n\leq\max\{1,N-2\}$.
\end{proof}

\begin{thm}\label{TIGHT1}
There exist instances which achieve the bound $n=\max\{1,N-2\}$
in Theorem \ref{Min3}.
\end{thm}
\begin{proof}
Fig. \ref{rgab} (a) is an instance of region graph with $N\geq 3$
sink regions and  $n=N-2$ coding regions, and (b) is an instance
of region graph with two sink regions and one coding regions. They
are both feasible because they have no singular region. We can
verify that these region graphs satisfy the two conditions of
Definition \ref{MFea}, hence are minimal feasible region graph.
\end{proof}

Now we can prove the main result of this section.

\begin{proof}[Proof of Theorem \ref{encn}]
Let $\mathcal G=(D,\mathcal E)$ be a minimal feasible region graph
of $G$ and $\tilde{C}$ be a linear code of $\mathcal G$. By Lemma
\ref{FES}, we can extend $\tilde{C}$ to a linear solution $C$ of
$G$, and $e\in E$ is an encoding link of $C$ only if $e$ is the
leader of some non-source region. Now, we prove that if $\mathcal
G$ is a minimal feasible region graph and $e$ is the leader of a
non-source region, then $e$ is an encoding link. Without loss of
generality, assume $e=\text{lead}(R)$. If $d_e=d_p$ for some
$p\in\text{In}(e)$, then by the construction of $C~($refer to
Lemma \ref{FES}$)$, $d_R=d_P$ for some $P\in\text{In}(R)$. Then we
can contract $D$ by combining $P$ and $R$ and obtain a feasible
region decomposition of $G$, which contradicts to the minimality
of $\mathcal G$. So $d_e\neq d_p$ for any $p\in\text{In}(e)$, and
$e$ is an encoding link of $C$. Thus, if $\mathcal G$ is a minimal
feasible region graph, then $e$ is an encoding link if and only if
$e$ is the leader of a non-source region.

By 1) of Corollary \ref{Min2}, a non-source region is either a
coding region or a sink region. By Remark \ref{num-sink}, the
number of sink regions is at most $N$. By Theorem \ref{Min3}, the
number of coding regions is at most $\max\{1,N-2\}$. Thus, the
number of encoding links is at most
$N+\max\{1,N-2\}=\max\{3,2N-2\}$.

Now, consider networks with minimal feasible region graph as in Fig.
\ref{rgab}. We have shown that they have $N$ sink regions and
$\max\{1,N-2\}$ coding regions $($refer to Theorem \ref{TIGHT1}$)$.
So the number of encoding links is $\max\{3,2N-2\}$.
\end{proof}

\section{Bound on Field Size}
In this section, following the same technique as in
\cite{Fragouli06} $($by converting a network coding problem to a
graph coloring problem \cite{Fragouli04}$)$, we derive an upper
bound on the field size for the 2-URMS problem. The result shows
that it is not necessary to use a larger field for 2-URMS network
than for one multicast session with two single rate flows
\cite{Fragouli06}.

First, we strengthen the conclusion of Lemma \ref{FRGC} under the
condition that $\mathcal G$ is a minimal feasible region graph.

\begin{lem}\label{FRGC-1}
Let $\mathcal G=(D,\mathcal E)$ be a minimal feasible region graph
of $G$. Then $\tilde{C}=\{d_{R}\in \mathbb F^2; R\in D\}$ is a
linear code of $\mathcal G$ if and only if the following two
conditions hold.
\begin{itemize}
    \item[(1)] If $R$ is an $X_i$ region, then
    $d_R=\alpha_i~(i\in\{1,2\})$, where $\alpha_1=(1,0)$ and $\alpha_2=(0,1)$;
    \item[(2)] If $\{R,Q\}\subseteq D$ such that $R,Q$ have a common child
    then $d_{R}$ and $d_{Q}$ are linearly independent.
\end{itemize}
\end{lem}
\begin{proof}
Suppose $\tilde{C}$ satisfies conditions (1) and (2). Then
clearly, $\tilde{C}$ satisfies the conditions of Lemma \ref{FRGC},
hence is a linear code of $\mathcal G$.

Conversely, suppose $\tilde{C}$ is a linear code of $\mathcal G$. By
Definition \ref{Fea}, condition (1) holds. For condition (2),
suppose $R,Q$ have a common child $W$. If $d_{R}$ and $d_{Q}$ are
linearly dependent, then by properly naming, $d_Q=\lambda d_R$ for
some $\lambda\in\mathcal F$. We delete the edge between $Q$ and $W$,
and obtain a region graph $\mathcal G'$. Clearly, $\tilde{C}$ is a
linear code of $\mathcal G'$, and $\mathcal G'$ is feasible, which
contradicts to the minimality of $\mathcal G$. So $d_{R}$ and
$d_{Q}$ are linearly independent. Hence condition (2) holds.
\end{proof}

If $\mathcal G=(D,\mathcal E)$ is a minimal feasible region graph
and $\{R,Q\}\subseteq D$ such that $R,Q$ have a common child, then
by 2) of Theorem \ref{Min}, either $R$ and $Q$ are both coding
region. Or, by proper naming, $R$ is a coding region and $Q$ is an
$X_j$ region $(j=1$ or $2)$.

To convert the network coding problem to a graph coloring problem,
we need first define the \emph{associated graph} of $\mathcal G$.

\begin{defn}\label{a-g}
Let $\mathcal G=(D,\mathcal E)$ be a minimal feasible region graph
of $G$ having $n$ coding regions $R_1,\cdots,R_n$. The associated
graph $\Omega_{\mathcal G}$ is defined as an undirected graph with
vertex set $V(\Omega_{\mathcal G})=\{X_1,X_2,R_1,\cdots,R_n\}$,
and its edge set includes the following:
\begin{enumerate}
  \item $(X_1,X_2)$: it is called the red edge of $\Omega_{\mathcal G}$.
  \item $(R_i,R_j)$: if $R_i$ and $R_j$ have a common child, it is
  called a blue edge of $\Omega_{\mathcal G}$;
  \item $(R_i,X_j)$: if $R_i$ have a common child with some $X_j$
  region $(j=1,2)$, it is called a green edge of $\Omega_{\mathcal G}$.
\end{enumerate}
\end{defn}

\begin{figure}[htbp]
\begin{center}
\includegraphics[width=5.5cm]{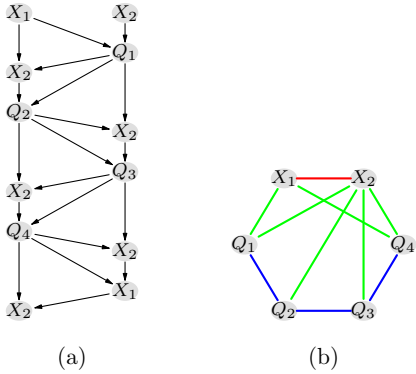}
\end{center}
\caption{(a) is a minimal feasible region graph; (b) is the
associated graph.} \label{agmg}
\end{figure}

Fig. \ref{agmg}(a) shows a minimal feasible region graph, which is
an instance of the minimal feasible region graph in Fig.
\ref{rgab}(a) with $n=4$. We depict its associated graph in Fig.
\ref{agmg}(b).

\renewcommand\figurename{Fig}

If $\tilde{C}=\{d_{R}; R\in D\}$ is a linear code of $\mathcal G$,
the same coloring arguments as in \cite{Fragouli06} can be applied
to color the graph $\Omega_{\mathcal G}$ with the mapping
$\rho:V(\Omega_{\mathcal
G})\rightarrow\{\alpha_1,\alpha_2,d_{R_1},\cdots,d_{R_n}\}$. Thus,
we can obtain the following lemma.

\begin{lem}\label{NVP}
Let $\mathcal G=(D,\mathcal E)$ be a minimal feasible region graph
of $G$. Then $\mathcal G$ has a linear code over the field of size
$q$ if and only if $q\geq\chi(\Omega_{\mathcal G})-1$, where
$\chi(\Omega_{\mathcal G})$ is the chromatic number of
$\Omega_{\mathcal G}$.
\end{lem}

Note that besides the source regions and coding regions, the
region graph of 2-URMS networks contains sink regions. So the
arguments in \cite{Fragouli06} can not be directly applied to
2-URMS problem. To derive our result, we need the following lemmas
to estimate the chromatic number of $\Omega_{\mathcal G}$.

\begin{lem}\label{CCTS}
Let $\mathcal G=(D,\mathcal E)$ be a minimal feasible region graph
of $G$ with $n~(n\geq 1)$ coding regions. Then the $X_1$ source
region and $X_2$ source region have a common child.
\end{lem}
\begin{proof}
Since $\mathcal G$ is acyclic, its regions can be ordered in an
upstream-to-downstream fashion, i.e., $\forall R,R'\in D$, $R<R'$
if $R$ is a parent of $R'$. Let $R$ be the least non-source
region, i.e., for any non-source region $R'\neq R$, $R<R'$. By 1)
of Theorem \ref{Min}, any non-source region has exactly two
parents in $\mathcal G$. Then $R$ is a common child of the $X_1$
source region and $X_2$ source region.
\end{proof}

\begin{lem}\label{CRCC}
Let $\mathcal G=(D,\mathcal E)$ be a minimal feasible region graph
of $G$ such that the number of coding regions is at least one.
Then every vertex in $\Omega_{\mathcal G}$ has degree at least
$2$.
\end{lem}
\begin{proof}
1) Vertices $X_1$ and $X_2$: Since $\mathcal G$ is acyclic and
finite, there must be a coding region, say $R$, such that no child
of $R$ is a coding region. By Definition \ref{a-g} and 3) of
Corollary \ref{Min2}, both $(R,X_1)$ and $(R,X_2)$ are edges of
$\Omega_{\mathcal G}$. Moreover, by Definition \ref{a-g},
$(X_1,X_2)$ is an edge of $\Omega_{\mathcal G}$. So the vertices
$X_1$ and $X_2$ both have degree at least $2$.

2) Coding regions: For any coding region $R$, we have the
following two cases.

Case 1: No child of $R$ is coding region. As proved in 1), both
$(R,X_1)$ and $(R,X_2)$ are edges of $\Omega_{\mathcal G}$. So $R$
has degree at least $2$.

Case 2: $R$ has a child $Q$ which is a coding region. By 3) of
Theorem \ref{Min}, $R$ and $Q$ have a common child. Hence $(R,Q)$
is an edge of $\Omega_{\mathcal G}$. Moreover, by 2) of Corollary
\ref{Min2}, $R$ has a child $W$ which is an $X_i$ sink region for
some $i\in\{1,2\}$. By 4) of Theorem \ref{Min}, $R$ has a common
child with an $X_i$ region. Hence $(R,X_i)$ is an edge of
$\Omega_{\mathcal G}$. So $R$ has degree at least $2$.

From all the above discussions, each vertex of $\Omega_{\mathcal
G}$ has degree at least $2$.
\end{proof}

\begin{lem}\label{Chr}
[\cite{JA79}, Ch. 9] Every $k$-chromatic graph has at least $k$
vertices of degree at least $k-1$.
\end{lem}

Now, we can present our main result of this section.

\begin{thm}\label{FSize}
Suppose $G$ is a solvable 2-URMS network with $N$ sinks. Then $G$
has a linear solution over the field with size no larger than
$\max\{2,\lfloor\sqrt{2N-7/4}+1/2\rfloor\}$.
\end{thm}
\begin{proof}
If $N=2$, a binary field is sufficient for a solution
\cite{Wang110}. We prove that the field of size
$\lfloor\sqrt{2N-7/4}+1/2\rfloor$ is sufficient for a solution when
$N\geq 3$.

Let $\mathcal G=(D,\mathcal E)$ be a minimal feasible region graph
of $G$ having $n$ coding regions and $K$ sink regions. Let $J$ be
the number of edges of the associated graph $\Omega_{\mathcal G}$,
and let $k=\chi(\Omega_{\mathcal G})$ be the chromatic number of
$\Omega_{\mathcal G}$. We count $J$ in two different ways.

By Lemma \ref{CRCC} and \ref{Chr}, each vertex of
$\Omega_{\mathcal G}$ has degree at least $2$ and at least $k$
vertices with degree at least $k-1$. We have \begin{equation}
[k(k-1)+2(n+2-k)]/2\leq J.\tag{5} \end{equation}

On the other hand, by 1) of Theorem \ref{Min}, a region is a common
child of two regions if and only if it is a non-source region.
Moreover, by 1) of Corollary \ref{Min2}, a non-source region is
either a coding region or a sink region. So there are $n+K$ regions
which are common children of two regions. By Lemma \ref{CCTS}, among
these $n+K$ regions, one of them is the common child of the $X_1$
source region and $X_2$ source region. So the total number of blue
edges and green edges of $\Omega_{\mathcal G}$ is at most $n+K-1$.
Plus the red edge, the total number of edges of $\Omega_{\mathcal
G}$ is at most $n+K$. That is, \begin{equation} J\leq n+K.\tag{6}
\end{equation}

Combining equations (5) and (6), we have:
$$[k(k-1)+2(n+2-k)]/2\leq n+K.$$ Then
$$k^2-3k+4\leq 2K.$$ By Remark \ref{num-sink}, $K\leq
N$. So \begin{equation} k^2-3k+4\leq 2N.\tag{7}
\end{equation} Solving equation (7), we have
$k\leq\sqrt{2N-7/4}+3/2$.

By Lemma \ref{NVP}, a field with size no larger than
$k-1=\sqrt{2N-7/4}+1/2$ is sufficient for a linear solution.
\end{proof}

In the following, we show that the bound in Theorem \ref{FSize} is
tight.

\begin{thm}\label{Bound}
There exist instances of networks for which the field size bound
in Theorem \ref{FSize} is achieved.
\end{thm}
\begin{proof}
We shall construct a minimal feasible region graph by adding some
$X_2$ sink regions to the region graph in Fig.\ref{rgab} (a). The
process is as follows.
\begin{enumerate}
  \item For $j\in\{2,\cdots,n-1\}$, add an $X_2$ sink region as
  a common child of $Q_j$ and the $X_1$ source region;
  \item For each pair $\{Q_i,Q_j\}$ such that $Q_i$ and $Q_j$ are
  not adjacent, add an $X_2$ sink region as a common child of $Q_i$ and $Q_j$.
\end{enumerate}
Denote the resulted region graph by $\mathcal G$ and the
corresponding vertex set by $D$. One can check that $\mathcal G$
is still a minimal feasible region graph and the associated graph
$\Omega_{\mathcal G}$ is a complete graph with $n+2$ vertices.

Clearly, $n-2$ sink regions are added in step 1). In addition,
note that there are $n-2$ coding regions not adjacent to $Q_1$,
$n-3$ coding regions not adjacent to each $Q_j~(j=2,\cdots,n-1)$,
and $n-2$ coding regions not adjacent to $Q_n$. Thus, there are
$[2(n-2)+(n-2)(n-3)]/2=(n^2-3n+2)/2$ sink regions be added in step
2). Plus the $n+2$ sink regions in the original graph, the total
number of sink regions of $\mathcal G$ is
$N=(n+2)+(n-2)+(n^2-3n+2)/2=(n^2+n+2)/2$. Solving $n$ from this
equation, we have $n=\sqrt{2N-7/4}-1/2$.

Since the associated graph $\Omega_{\mathcal G}$ is a complete
graph with $n+2$ vertices, the chromatic number of
$\Omega_{\mathcal G}$ is $\chi(\Omega_{\mathcal G})=n+2$. By Lemma
\ref{NVP}, the field size for any linear code of $\mathcal G$ is
at least $n+1=\sqrt{2N-7/4}+1/2$.
\end{proof}

\renewcommand\figurename{Fig}
\begin{figure}[htbp]
\begin{center}
\includegraphics[height=4.3cm]{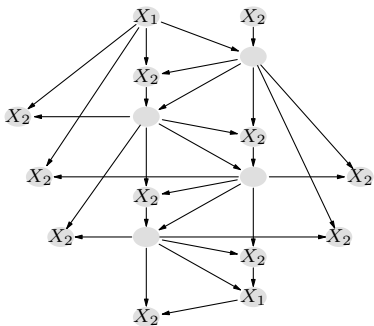}
\end{center}
\caption{A minimal feasible region graph with $n=4$ coding regions
and $N=(n^2+n+2)/2=11$ sink regions.} \label{emgb}
\end{figure}

\begin{exam}
Fig. \ref{emgb} plots a region graph $\mathcal G$ constructed as
in the proof of Theorem \ref{Bound}. The graph $\mathcal G$ has
$n=4$ coding regions and $N=(n^2+n+2)/2=11$ sink regions. By
Theorem \ref{FSize} and \ref{Bound}, $\mathbb F_{5}$ is the
smallest field that can ensure a linear solution of $\mathcal G$.
\end{exam}

\section{Conclusions and Discussions}
We investigated the encoding complexity of network coding with two
unit-rate multicast sessions $($2-URMS$)$ by a region
decomposition method. We showed that when the network is
decomposed into mutually disjointed regions, a network coding
solution can be easily obtained from some simple labelling
operations on the region graph and through assignment of
decentralized coding vectors. All the processes of the region
decomposition, the region labelling, and the code construction can
be done in time $O(|E|)$.

We further reduced a feasible region graph into a minimal feasible
region graph by deleting edges and/or combining nodes of the
region graph. We showed that the minimal feasible region graph
have some interesting local properties, from which we derived
bounds on the encoding links and on the required field size.

In this paper, we only consider scalar network coding, and the
field size bound in Theorem \ref{FSize} is shown to be tight. But
for vector network coding, as pointed out in \cite{Ebrahimi11}, it
is possible to use smaller field while increase the vector length,
which serves as an interesting future work.




\section*{Acknowledgment}
This research is partly supported by NSF of China (No. 10990011),
the research fund for the Doctoral Program of Higher Education of
China, and the International Design Center (grant no. IDG31100102
\& IDD11100101). The work of Kai Cai is partially supported by a
grant from the University Grants Committee of the Hong Kong
Special Administrative Region, China (Project No.
AoE/E-02/08).\ifCLASSOPTIONcaptionsoff
  \newpage
\fi

\begin{IEEEbiographynophoto}{Wentu Song}
received the BS and MS degrees in Mathematics from Jilin
University, China in 1998 and 2006, respectively, and the Ph.D.
degree in Mathematics from Peking University in 2012, China. He is
currently a Post Doc research fellow in Singapore University of
Technology and Design, where he is working on network coding and
network distributed storage.
\end{IEEEbiographynophoto}

\begin{IEEEbiographynophoto}{Kai Cai}
received the M.S. and Ph.D. degrees in mathematics from Peking
University, Beijing, China, in 2001 and 2004, respectively. From
2004 to 2006, he was a postdoc research fellow of the Electronic
Engineering Department, Tsinghua University, Beijing, China. From
2006 to 2007, he visited the department of Electronic and Computer
Engineer, the Hong Kong University of Science and Technology. From
2007 to 2011, he was an associate professor in the Institute of
Computing Technology, Chinese Academy of Sciences. He visited
School of Electrical, Computer and Energy Engineering, Arizona
State University from 2011 to 2012. He is now with Institute
of Network Coding, The Chinese University of Hong Kong. His
research interests are in algebraic combinatorics, coding theory,
network information theory, and so forth.
\end{IEEEbiographynophoto}

\begin{IEEEbiographynophoto}{Rongquan Feng}
received the Ph.D. in Mathematics from the Institute of Systems
Science, Chinese Academy of Sciences in 1994. He is currently a
professor in Peking University. His research interests are in the
areas of algebraic combinatorics, cryptology, coding theory and
information security. He served as the secretary-general of
Beijing Mathematical Society from 2005. He is an associate editor
of the journal Mathematics in Practice and Theory.
\end{IEEEbiographynophoto}

\begin{IEEEbiographynophoto}{Chau Yuen}
received the B. Eng and PhD degree from Nanyang Technological
University, Singapore in 2000 and 2004 respectively. Dr Yuen was a
Post Doc Fellow in Lucent Technologies Bell Labs, Murray Hill
during 2005. He was a Visiting Assistant Professor of Hong Kong
Polytechnic University in 2008. During the period of 2006 2010, he
worked at the Institute for Infocomm Research (Singapore) as a
Senior Research Engineer. He joined Singapore University of
Technology and Design as an assistant professor from June 2010. He
serves as an Associate Editor for IEEE Transactions on Vehicular
Technology. On 2012, he received IEEE Asia-Pacific Outstanding
Young Researcher Award.
\end{IEEEbiographynophoto}




\vfill



\end{document}